\documentclass[prb,amssymb,twocolumn,showpacs]{revtex4}
\usepackage{bm}
\usepackage{graphicx}
\usepackage{amsmath}
\usepackage{color}

\begin{document}

\title{Electron quantum dynamics in closed and open 
potentials at high magnetic fields: Quantization and lifetime effects
unified by semicoherent states}

\author{Thierry Champel}
\affiliation{Laboratoire de Physique et Mod\'{e}lisation des Milieux Condens\'{e}s, CNRS
and Universit\'{e} Joseph Fourier, B.P. 166,
25 Avenue des Martyrs, 38042 Grenoble Cedex 9, France}

\author{Serge Florens}
\affiliation{Institut N\'{e}el, CNRS and Universit\'{e} Joseph Fourier,
B.P. 166, 25 Avenue des Martyrs, 38042 Grenoble Cedex 9, France }

\date{\today}

\begin{abstract}
We have developed a Green's function formalism based on the use of an
overcomplete semicoherent basis of vortex states specially devoted to the study
of the Hamiltonian quantum dynamics of electrons at high magnetic fields
and in an arbitrary potential landscape smooth on the scale of the magnetic
length. This formalism is used here to derive the exact Green's function for
an arbitrary quadratic potential in the special limit where Landau level
mixing becomes negligible. This solution remarkably embraces under a unified form
the cases of confining and unconfining quadratic potentials. This property results from the
fact that the overcomplete vortex representation provides a more general type of
spectral decomposition of the Hamiltonian operator than usually considered.
Whereas confining potentials are naturally characterized by quantization
effects, lifetime effects emerge instead in the case of saddle-point potentials.
Our derivation proves that the appearance of lifetimes has for origin the instability
of the dynamics due to quantum tunneling at saddle points of the potential landscape.
In fact, the overcompleteness of the vortex representation reveals an intrinsic
microscopic irreversibility of the states synonymous with a spontaneous breaking of
the time symmetry exhibited by the Hamiltonian dynamics.

\end{abstract}


\maketitle

\section{Introduction}

The integer quantum Hall effect, with its remarkable transport properties, 
\cite{Klitzing1980} offers perhaps the simplest route to understand the
complex dynamics of electrons taking place in strongly inhomogeneous
nanostructures.
Indeed, the presence of a strong perpendicular magnetic field in two dimensions
brings the classical motion close to integrability, with slow drift trajectories
superimposed to faster cyclotron orbits, leaving hope that the quantum dynamics displays
similar and simple structures. \cite{Iordansky1982,Kazarinov1982,Trugman1983,Apenko1983,
Joynt1984,Girvin1984,Shapiro1986,Prange1987} Further simplification is brought
by the fact that electron-electron interactions can be taken into account in the
integer quantum Hall regime at the single-particle level,
\cite{Chklovskii1992,Klitzing2005} so that the calculation of equilibrium properties,
such as the local electron density and the distribution of permanent currents
throughout the sample, can be carried out from a one-particle random
Schr\"{o}dinger equation. Nevertheless, the precise microscopic resolution of
this equilibrium one-body problem still lacks a concrete analytical formalism and
progress in this direction would be useful toward a microscopic description that
underlies more complex nonequilibrium phenomena.

The main difficulty in the resolution of the disordered Schr\"{o}dinger equation
resides in the complexity of the potential landscape that competes with the
kinetic energy in a nontrivial manner. In fact, it is worth mentioning that the
standard procedure \cite{Abrikosov} to deal with a random potential, which
consists in averaging over impurity configurations,
is already physically questionable at high magnetic fields, at least at the microscopic
level. Indeed, this theoretical route is usually justified by the physical assumption of
randomness after successive collision events. However, instead of the chaotic
exploration of the disordered landscape, the electronic classical motion becomes
relatively regular for a smooth disorder potential at high magnetic fields. At the
technical level, this difficulty was pointed out in the standard
quantum-mechanical diagrammatic perturbative method, which leads to unsolved
complications for a smoothly disordered potential: more and more classes of
diagrams must be incorporated in the calculation as the strength of the
magnetic field is increased. \cite{Raikh1993} This somehow indicates that the
perturbative technique is unadapted for the high magnetic-field regime when the
bending of the electronic trajectories becomes too important.

We note already that if one does not resort to any averaging over disorder, one faces not only
a technical problem but also a more fundamental physical one, namely, the question of the
microscopic origin of irreversibility and intrinsic dissipation from Hamiltonian
quantum dynamics, an essential aspect for the calculation of transport properties.
Indeed, the standard impurity averaging procedure, valid in the limit of low
magnetic fields, introduces an effective description of the formalism. This allows to get,
in addition to the energy spectrum, lifetime effects, which are inaccessible in a
purely Hamiltonian formalism limited to the Hilbert space of square integrable functions.
In absence of averaging, it is thus necessary to clarify the possible relation between
the ``complexity'' of the potential landscape and the issue on the microscopic origin of
irreversible processes. In other words, we are confronted with the controversial question
whether irreversibility results from supplementary approximations to the
fundamental quantum-mechanical laws (which are strictly time reversible) or
is subtly hidden in the usual formulation of quantum dynamics.

A well-known starting point to capture the regime of quasi-regular dynamics
in high perpendicular magnetic fields is to implement in full exact quantum-mechanical terms the fast cyclotron motion of circling electrons resulting 
from the Lorentz force, which gives rise to the quantization of the kinetic 
energy into discrete Landau levels. The second relevant degree of freedom 
then corresponds to the slow guiding center of motion, whose dynamics is
dictated by the smooth disordered potential landscape. The essential role of
the potential landscape, to be captured precisely in the microscopic quantum-mechanical calculations, is to lift the huge degeneracy of the Landau levels.
It is already important to note that the magnetic field $B$ enters into the quantum-mechanical problem only via two different quantities: the cyclotron pulsation
$\omega_{c}=|e|B/m^{\ast}c$ and the magnetic length $l_{B}=\sqrt{\hbar c/|e|B}$.
While $\omega_{c}$ is a material-dependent parameter via the effective mass
$m^{\ast}$, $l_{B}$ can be regarded as a more fundamental quantity since it
involves only physical constants such as Planck's constant $\hbar$, the speed of
light $c$, and the absolute value of electric charge $|e|$. The cyclotron
pulsation $\omega_{c}$ determines a characteristic frequency for the circular
motion which actually already enters into the problem at the classical level. In
contrast, the magnetic length $l_{B}$ is purely a quantum-mechanical quantity,
especially characterizing the spatial extent of the wave functions. In the popular
operatorial language of quantum mechanics, these two relevant degrees of freedom are introduced by decomposing the
electronic coordinate $\hat{{\bf r}}=\hat{{\bf R}}+\hat{{\bm \eta}}$ into a
relative position $\hat{{\bm \eta}}=\hat{{\bf v}} \times \hat{z} /\omega_{c}$
linked to quasicircular cyclotron orbits ($\hat{{\bf v}}$ is the velocity operator), and
a guiding center position $\hat{{\bf R}}=(\hat{X},\hat{Y})$. These quantum variables obey
the commutation rules $[\hat{v}_{x},\hat{v}_{y}]=-i\hbar \omega_c/m^\ast$ and $[\hat{X},\hat{Y}]=i
l_{B}^{2}$. In close analogy with the canonical quantization rule between the
position and the momentum, it is seen that, for the slow drift motion of the
guiding center, $l_{B}^{2}$ plays the role of an effective magnetic-field-dependent Planck's constant.

The condition of a high magnetic field can thus be imposed either by expressing that
$l_{B}$ is the smallest length scale in the problem, i.e., by taking $l_{B} \to
0$, or by considering that $\omega_{c}$ is the biggest frequency scale in the
problem, i.e., by taking $\omega_{c} \to \infty$. These two limits can in
principle be taken separately or simultaneously and actually yield different
physical situations which have been discussed in the literature.
\cite{Trugman1983,Apenko1983,Joynt1984,Girvin1984,Shapiro1986,Jain1987,Jain1988a,Jain1988b,Entelis1992,Geller1994,Rohringer2003}
For instance, a popular approach corresponding to the first limit $l_{B} \to 0$
is to treat classically the slow guiding center motion while the fast cyclotron
motion is kept quantum mechanical.
\cite{Trugman1983,Apenko1983,Joynt1984,Shapiro1986,Geller1994}
This case leads to great simplifications in the theoretical treatment, since the
guiding center coordinates then commute and can be described entirely in classical
terms. In this limit, the guiding center motion is restricted to equipotential lines
and the energy spectrum is characterized by a continuous potential energy on top of
discrete Landau levels.
Another standard approximation corresponding to the second limit $\omega_{c} \to \infty$
is to neglect Landau level mixing and to study the Hamiltonian quantum dynamics
at finite $l_B$ projected onto a single Landau level.
\cite{Girvin1984,Jain1987,Jain1988a,Jain1988b,Entelis1992,Rohringer2003}
This most tricky regime implies to work in a fully quantum-mechanical formalism taking
into account rigorously the noncommutativity of the guiding center coordinates.
An interesting aspect that was not fully cleared with both types of approaches
lies in how to capture the transition from quantum to classical,
with the classical features emerging possibly from microscopic decoherence
processes.

In order to study the interplay of Landau levels quantization and a smooth
disordered potential in a controlled fashion, we have developed
\cite{Champel2007,Champel2008} in recent years a specially devoted Green's
function formalism based on the use of a semicoherent overcomplete set of
states $|m,{\bf R} \rangle $ labeled by a continuous quantum number ${\bf
R}$, related to the classical guiding center coordinates, and an integer $m$,
associated to the discrete Landau levels. This family of states were named vortex
states \cite{Champel2007} due to the vortexlike phase singularity of the
associated wave functions $\langle {\bf r} | m,{\bf R} \rangle$ at the electronic
position ${\bf r}={\bf R}$. Because the vortex states encode no preferred symmetry, they
allow a great adaptability to the local spatial variations of the random
potential. More precisely, our approach consists in mapping the quantum
equation of motion obeyed by the Green's function (the so-called Dyson equation)
to this vortex representation, which then rigorously extends to quantum mechanics
the classical guiding center picture. The main essential difference with the guiding center treatment
is that our method keeps the full quantum-mechanical noncommutativity of the
guiding center coordinates through the overcompleteness property of the basis of
states. We have shown \cite{Champel2008} that, within the vortex representation,
Dyson equation can be easily and systematically diagonalized order by order in
powers of the magnetic length $l_{B}$. Quantum observables are then obtained by
returning to electronic representation from the vortex Green's functions, so
that the semiclassical limit $l_{B} \to 0$ as well as its systematic corrections is
naturally obtained with our approach. \cite{Champel2008} Moreover, the vortex
representation allows one to classify and include in a systematic and straightforward
way the Landau level mixing processes in the calculations.

The results to be developed here aim at extending the work initiated in
Refs.~\onlinecite{Champel2007,Champel2008}, where a systematic and closed form
expression for the solution of Dyson equation in a smooth arbitrary potential
was already obtained, under the form of a series expansion classified order by order in
powers of $l_{B}$.
The further and important step made in the present paper is to include to all orders the
contributions from first and second spatial derivatives of the potential,
in loose analogy to the resummation of leading classes of Feynman diagrams in
standard perturbation theory. For simplification, we will consider the mathematical
limit where Landau level mixing can be considered negligible and our solution
will encompass all cases of quadratic potentials in that limit.
A further motivation for a resummation of the gradient expansion is that the series
obtained in powers of $l_{B}$ may not converge in general, since the semiclassical
guiding center limit $l_{B} \to 0$ is expected to be singular, similar to the case of
the more standard, fully semiclassical limit $\hbar \to 0$. That the small $l_B$ expansion
is indeed singular will be illustrated by the asymptotic character of guiding center
semiclassical results: physical aspects related to an exact quantum treatment, such as
quantization of energy levels or lifetime effects, can not be approximated by a finite
expansion. We shall confirm this feature by comparing our exact quantum solution
to various approximation schemes related to several improvements of the semiclassical guiding
center method.

Before diving into the heart of the technique, we want to mention that, although
our method obviously shares on certain aspects some similarities with theories
already existing in the literature, important differences with these prior works
can be emphasized. First, we would like to stress that our methodology
is based on the exclusive use of Green's functions, not wave functions, in contrast to the
theory pioneered by Girvin and Jach \cite{Girvin1984} (see also Refs.
\onlinecite{Jain1988a,Entelis1992,Rohringer2003}) where a one-dimensional (1D)
Schr\"{o}dinger's equation for the electron dynamics projected onto a single Landau
level (valid in the limit $\omega_{c} \to \infty$) was derived for finite $l_{B}$.
This point could appear superfluous at the first glance, since it is
possible to get Green's functions from the knowledge of wave functions. However,
the use of an overcomplete representation with nonorthogonal states to solve the dynamical equations of
motion necessarily forces us to give up the wave functions picture and work in
a Green's function formalism of partially coherent states. Furthermore, it is worth
noting that the Hilbert space of square integrable wave functions is usually well
suited for closed integrable systems, but turns out to be totally inadequate in
situations presenting scattering processes in open systems (case of a saddle-point
potential for instance), where one must appeal to another formalism, the scattering
states picture. We shall show that the use of an overcomplete representation of
coherent states allows one to get and treat quantization and lifetime
effects on an equal footing in the resolution of Dyson equation. Moreover, the
appearance of lifetimes in the energy spectrum coincides with the impossibility
to describe the solution of the Dyson equation in terms of a countable set of states, thus
proving the relevance of an overcomplete \cite{note} representation in such a situation.
Second, we note that several authors \cite{Girvin1984,Jain1987,Jain1988a,Jain1988b}
actually attempted to build theories based on the use of vortex states within
the path-integral formalism, which, however, seemed to suffer from
technical difficulties that were not elucidated.
\cite{Girvin1984,Jain1987,Jain1988a} In contrast, our theory is not tainted with
the specific mathematical ambiguities which can often be encountered with the
path-integral technique.

The derivation of an exact solution for the Green's function at large cyclotron
energy yet finite magnetic length, embracing all possible cases of quadratic
potentials, constitutes the main mathematical result of this paper.
Besides capturing exactly the tunneling processes in the case of
a saddle-point potential, it has the virtue of pointing out clearly the
physical microscopic mechanism responsible for the appearance of lifetimes in
the spectral decomposition of the Hamiltonian. We therefore hope that it will
also help to clarify the debate about the physical roots of time irreversibility
and the ubiquitous emergence of a classical character from quantum mechanics.
An important point we will also demonstrate is that the derived solution provides
a controlled approximation at finite temperature for all equilibrium local observables in the case of an
{\it arbitrary} potential that is smooth on the scale of the magnetic length.
This result, based on the fact that the local Green's function at high magnetic
fields displays a hierarchy of energy scales controlled by successive spatial
derivatives of the potential, can be used to write down an expression for the local
density of states which may be useful in the context of recent scanning tunneling
spectroscopy measurements. \cite{Hashimoto2008} A short report of parts of this
work has been published in Ref. \onlinecite{Champel2009}.

The paper is organized as follows. In Sec. \ref{Section2} we present the vortex
Green's function formalism and derive the general form of Dyson equation in the vortex
representation. In Sec. \ref{Section3}, Dyson equation is exactly solved for the
two particular cases of an arbitrary 1D potential and an arbitrary two-dimensional (2D) quadratic
potential. The obtained solutions are then exploited in Sec. \ref{Section4} to
derive a general expression for the local density of states. Finally, we discuss
in Sec. \ref{Section5} the importance of considering an overcomplete
representation in the present problem and its physical implications for the
issue of time irreversibility. A small conclusion closes the paper. Some extra
technical details bringing complements of information for the calculations are
given in several appendixes.

\section{Dyson equation in vortex space}
\label{Section2}

\subsection{Hamiltonian and projection onto the vortex representation}

We consider the single-particle Hamiltonian for an electron of charge $e=-|e|$
confined to a two-dimensional $(x,y)$ plane in the presence of both a
perpendicular magnetic field ${\bf B}$ and an arbitrary potential energy $V({\bf r})$,
\begin{equation}
H=\frac{1}{2 m^{\ast}} \left(-i \hbar {\bm \nabla}_{{\bf r}}-\frac{e}{c}{\bf A}({\bf r}) \right)^{2}+V({\bf r}),
\label{Ham}
\end{equation}
with the vector potential ${\bf A}$ defined by ${\bm \nabla} \times {\bf A}={\bf
B}=B \hat{{\bf z}}$, and $m^{\ast}$ the electron effective mass (here ${\bf
r}=(x,y)$ is the position of the electron in the plane).

For $V=0$, the energy spectrum is quantized into Landau levels
$E_{m}=(m+1/2)\hbar \omega_{c}$ with $\omega_{c}=|e|B/(m^{\ast}c)$. The high
degeneracy of the energy levels in absence of a potential is associated with a
great freedom in the choice of a basis of eigenstates for the free Hamiltonian.
To diagonalize Hamiltonian (\ref{Ham}) for an arbitrary potential energy
landscape $V({\bf r})$, a very convenient basis \cite{Champel2007} turns out
to be the overcomplete set of so-called vortex wave functions given by
\begin{eqnarray}
\Psi_{m,{\bf R}}({\bf r}) &=&\langle {\bf r} | m,{\bf R} \rangle \\
&=&\frac{1}{\sqrt{2 \pi l_{B}^{2} m!}} \left(
\frac{z-Z}{\sqrt{2}l_{B}}\right)^{m}
e^{- \frac{|z|^{2}+|Z|^{2}-2Zz^{\ast}}{4 l_{B}^{2}}}, \hspace*{0.5cm}
\label{vortex}
\end{eqnarray}
with $z=x+iy$ and $Z=X+iY$. The continuous variable ${\bf R}=(X,Y)$ constitutes
the quantum analog of the semiclassical guiding center discussed in the
introduction.
Here we have expressed the wave functions (\ref{vortex}) in the symmetrical
gauge ${\bf A}={\bf B} \times {\bf r}/2$. Besides being eigenstates of the free
Hamiltonian, the set of wave functions (\ref{vortex}) has the coherent states
character with respect to the continuous (degeneracy) quantum number ${\bf R}$,
which also corresponds to a ``vortex''-like singularity for ${\bf
r}={\bf R}$. Despite being semiorthogonal, the set of quantum numbers $|m,{\bf
R} \rangle $ obeys the completeness relation
\begin{eqnarray}
\int \!\!\! \frac{d^{2} {\bf R}}{2 \pi l_{B}^{2}} \sum_{m=0}^{+ \infty} | m,{\bf R} \rangle \langle m,{\bf R} |=1,
\label{comp}
\end{eqnarray}
thus allowing one \cite{Champel2007} to use the vortex representation in a
Green's function formalism, providing unicity of the development, related to
the analyticity of the disorder potential. \cite{Glauber} Note that, however,
the nonorthogonality of the states prevents to build a perturbation theory
solely based on wave functions to deal with the potential term $V({\bf r})$.

We have shown in a previous work \cite{Champel2008} that the electronic Green's
function associated with Hamiltonian (\ref{Ham}) and satisfying the
evolution equation in the energy ($\omega$) representation (we set from now on
$\hbar =1$)
\begin{eqnarray}
\left(\omega - H \pm i \delta \right)G^{R,A}({\bf r},{\bf r}',\omega)=\delta({\bf r}-{\bf r}')
\label{Greenevol}
\end{eqnarray}
 can be written exactly in terms of vortex wave functions $\Psi_{m,{\bf R}}({\bf r})$ as
\begin{eqnarray}
G^{R,A}({\bf r},{\bf r}',\omega)= \int \!\! \frac{d^{2} {\bf R}}{2 \pi l_{B}^{2}}
\sum_{m,m'}^{+ \infty} \Psi_{m',{\bf R}}^{\ast}({\bf r}')\Psi_{m,{\bf R}}({\bf r}) \nonumber \\
\times
\sum_{p=0}^{+\infty}
\frac{1}{p!}\left( - \frac{l_{B}^{2}}{2}\Delta_{{\bf R}}\right)^{p}
g_{m;m'}^{R,A}({\bf R},\omega)
\label{Gelec}
.
\end{eqnarray}
Here $\Delta_{{\bf R}}$ means the Laplacian operator taken with respect to the
vortex position ${\bf R}$, and the term $\delta$ in the left-hand side of Eq. (\ref{Greenevol}) is an
infinitesimal positive quantity encoding the boundary condition for the time
evolution. The retarded Green's function $G^{R}$ (with plus sign in Eq. (\ref{Greenevol})) represents
the response of the system to an impulse excitation, while the advanced Green's
function $G^{A}$ (with minus sign) corresponds to a source wave with a
deltalike response. Note that the correspondence between Green's functions in
Eq.~(\ref{Gelec}) is nonlocal with respect to the Landau level index $m$,
as expected, but quasilocal with respect to the vortex position ${\bf R}$.

Equation (\ref{Greenevol}) for the electronic Green's function then maps
\cite{Champel2008} exactly onto the following Dyson equation for the vortex
Green's function $g_{m;m'}({\bf R},\omega)$ (from now on, we do not specify
that the Green's function depends on $\omega$ in order not to burden the
expressions):
\begin{eqnarray}
\left(
\omega -E_{m} \pm i \delta
\right)
g_{m;m'}^{R,A}({\bf R})
=
\delta_{m,m'}+ \sum_{m''=0}^{+\infty}\sum_{k=0}^{+ \infty} \left( \frac{l_{B}}{\sqrt{2}} \right)^{2k}
\nonumber \\
\times \frac{1}{k!}
\left( \partial_{X}-i\partial_{Y}\right)^{k} v_{m;m''}({\bf R})
\left( \partial_{X}+i\partial_{Y}\right)^{k} g_{m'';m'}^{R,A}({\bf R}).
\label{Dyson0}
\end{eqnarray}
The matrix elements $v_{m;m'}({\bf R})$ of the potential $V$ in the vortex
representation can be evaluated exactly for an arbitrary potential provided that
the latter is smooth, i.e., infinitely differentiable, which is the case for any
physical potential. They take the form of a series expansion in powers of the
magnetic length (see Ref. \onlinecite{Champel2007} for the detail of the
derivation)
\begin{eqnarray}
v_{m;m'}({\bf R})&=& \sum_{j=0}^{+ \infty} \left(\frac{l_{B}}{\sqrt{2}} \right)^{j}
v^{(j)}_{m;m'}({\bf R}) , \label{seriesv}\\
v_{m;m'}^{(j)}({\bf R})&=& \sum_{k=0}^{j}
\frac{(m+k)!}{\sqrt{m!\, m'!}} \frac{\delta_{m+k,m'+j-k}}{k!(j-k)!} \nonumber \\
&&
\times
(\partial_{X}+i\partial_{Y})^{k}
(\partial_{X}-i\partial_{Y})^{j-k} V({\bf R}).
\end{eqnarray}

\subsection{Systematic magnetic length expansion}
\label{section2b}
 One method adopted in the paper \cite{Champel2008} in order to solve Eq.
(\ref{Dyson0}) is to search the function $g_{m}^{R,A}({\bf R})$ under the form
of a series in powers of the magnetic length $l_{B}$, similarly to the matrix
elements of the potential,
\begin{equation}
g_{m;m'}^{R,A}({\bf R})=\sum_{j=0}^{+ \infty} \left(\frac{l_{B}}{\sqrt{2}} \right)^{j} g_{m ; m'}^{R,A \, (j)}({\bf R}).
\label{series}
\end{equation}
The functions $g_{m;m'}^{(j)}({\bf R})$ are then obtained by solving Eq.
(\ref{Dyson0}) order by order in powers of $l_{B}$. This leads to a
closed-recursive relation \cite{Champel2008} for the functions
$g_{m;m'}^{(j)}({\bf R})$, which allows one in principle to obtain an explicit
expression for $g_{m;m'}^{(j)}$ at any order $j$ from the knowledge of all other
components with subleading order $i<j$. The leading order component can be
readily obtained and reads
\begin{equation}
g_{m;m'}^{R,A \, (0)}({\bf R})=
\frac{\delta_{m,m'}}{\omega-E_{m}-V({\bf R})\pm i \delta}
\label{g0}
.
\end{equation}
Inserting expression (\ref{g0}) in Eq. (\ref{Gelec}) and keeping only the
$l_{B}$ zeroth order term coming with $p=0$ yields the compact expression
for the electronic Green's function
\begin{eqnarray}
G^{R,A\,(0)}({\bf r},{\bf r}',\omega)= \int \!\! \frac{d^{2} {\bf R}}{2 \pi l_{B}^{2}} \sum_{m=0}^{+ \infty} \Psi_{m,{\bf R}}^{\ast}({\bf r}')\Psi_{m,{\bf R}}({\bf r}) \nonumber \\
\times
g^{R,A \, (0)}_{m ;m}({\bf R})
\label{Gelec0}
,
\end{eqnarray}
which is a quite simple and general functional of $V({\bf R})$. Subleading
corrections up to order $l_B^3$ were explicitly calculated in Ref. \onlinecite{Champel2008}.

\subsection{Limitations of the strict $l_B$ expansion}
\label{section2c}
Because the term $v_{m;m'}^{(0)}({\bf R})= V({\bf R})\, \delta_{m,m'}$ in the series expansion
(\ref{seriesv}) is the dominant one for the matrix elements of a smooth potential
with characteristic length scale $\xi \gg l_{B}$, one could naively expect that
the leading component $g^{(0)}$ in Eq. (\ref{g0}) is also
the dominant one in the expansion (\ref{series}) for the Green's function.
As noted in Ref. \onlinecite{Champel2008},
this conclusion has however to be contrasted since it does not take into
consideration the fact that the $(l_{B}/\xi)$ expansion generates at higher
orders systematic terms which may be highly singular in energy due to their multiple
pole structure. This is most clearly seen from Eq.~(\ref{Gelec}) for the
electronic Green's function at coinciding points ${\bf r}={\bf r}'$ obtained
with the leading order vortex propagator $g^{(0)}$:
\begin{eqnarray}
\label{Gcomp}
G^{R,A}({\bf r},{\bf r},\omega)
\hspace{6.2cm}\\
\nonumber
=\sum_{m,p} \int \!\! \frac{d^{2} {\bf R}}{2 \pi l_{B}^{2}}
|\Psi_{m,{\bf R}}({\bf r})|^2
\frac{1}{p!}\left( - \frac{l_{B}^{2}}{2}\Delta_{{\bf R}}\right)^{p}
\frac{1}{\omega_m-V({\bf R})} \\
\nonumber
=\sum_{m,p} \int \!\! \frac{d^{2} {\bf R}}{2 \pi l_{B}^{2}}
\frac{1}{\omega_m-V({\bf R})}
\frac{1}{p!}\left( - \frac{l_{B}^{2}}{2}\Delta_{{\bf R}}\right)^{p}
|\Psi_{m,{\bf R}}({\bf r})|^2 \\
\nonumber
=\sum_{m,p}
\frac{1}{p!}\left( - \frac{l_{B}^{2}}{2}\Delta_{{\bf r}}\right)^{p}
\int \!\! \frac{d^{2} {\bf R}}{2 \pi l_{B}^{2}}
\frac{1}{\omega_m-V({\bf R})}
|\Psi_{m,{\bf R}}({\bf r})|^2,
\end{eqnarray}
where integrations by parts and the property
$|\Psi_{m,{\bf R}}({\bf r})|^2=|\Psi_{m,{\bf r}}({\bf R})|^2$ were used to
get the last line of Eq. (\ref{Gcomp}) (we have noted above $\omega_m=\omega-E_m\pm i\delta$).

Now clearly the truncation of the above Eq. (\ref{Gcomp}) to the first $p=0$ term
is only vindicated provided the integral varies on length scales larger than
$l_B$, which is not always guaranteed, as the vortex wave function spatially
extends precisely on the scale $l_B$. If these corrections become important, not
only the whole sum over $p$ above must be kept, but also all terms of similar
form that appear within the complete vortex Green's function $g_{m;m'}$ (i.e.,
to all orders in $l_B$). Let us see what kind of terms one should
then consider. By inspecting the second line in~(\ref{Gcomp}), one is in fact looking
for corrections in the vortex Green's function at order $l_B^2$ of the
type
\begin{equation}
\label{Gcorrection}
l_B^2 \Delta_{\bf R} \frac{1}{\omega_m-V({\bf R})}
= \frac{2 l_B^2 |\nabla_{\bf R} V|^2}{(\omega_m-V({\bf R}))^{3}}+
\frac{l_B^2 \Delta_{\bf R} V}{(\omega_m-V({\bf R}))^{2}}.
\end{equation}
Such terms with multiple poles, which indeed start to appear in $g^{(2)}$
(see Ref.~\onlinecite{Champel2008} for a complete derivation), proliferate
at all orders of the $l_B$ expansion, similar to the further contributions
associated to values of $p>1$ in Eq.~(\ref{Gcomp}). These corrections to the Green's
function will not be perturbatively small whenever one probes energies or temperatures
smaller than the first characteristic energy scale appearing above, namely,
$l_B|\nabla_{\bf R}V|$, in which case the leading expression~(\ref{Gelec0})
breaks down. Equation~(\ref{Gcorrection}) is however hinting at how a controlled
calculation can be performed: provided that a hierarchy of energy
scales $l_B|\nabla_{\bf R}V|\gg l_B^2\Delta_{\bf R}V\gg\ldots$ can be
established, a systematic resummation to all orders in $l_B$ of potential gradient
terms will push the validity of the calculation down to the smaller scale
$l_B^2\Delta_{\bf R}V$, and so on and so forth. This idea is quite analogous to
the usual resummation of classes of Feynman graphs in standard perturbation
theory and constitutes the basic motivation for the computations that will follow.

\subsection{Dyson equation in the absence of Landau level mixing}
To present the method of resolution of Dyson equation (\ref{Dyson0}) to infinite
order in the $l_{B}$ expansion, we shall focus for simplicity on the limit of vanishing
Landau level mixing, i.e, $\omega_{c} \to \infty$ with $l_B$ finite.
In this case, one can easily check that the vortex Green's function becomes
purely diagonal, $g_{m;m'}({\bf R})=g_{m}({\bf R})\delta_{m,m'}$, so that
Eq. (\ref{Dyson0}) gets simplified into
\begin{eqnarray}
\left(
\omega -E_{m} \pm i \delta
\right)
g_{m}^{R,A}({\bf R})
=
1+ \sum_{k=0}^{+ \infty} \left( \frac{l_{B}}{\sqrt{2}} \right)^{2k}
\frac{1}{k!}
\nonumber \\
\times
\left( \partial_{X}-i\partial_{Y}\right)^{k} v_{m}({\bf R})
\left( \partial_{X}+i\partial_{Y}\right)^{k} g_{m}^{R,A}({\bf R}),
\label{Dyson}
\end{eqnarray}
with
\begin{eqnarray}
v_{m}({\bf R})
=\sum_{j=0}^{+\infty} \frac{(m+j)!}{m!(j!)^{2}}
\left(\frac{l_{B}^{2}}{2} \Delta_{{\bf R}}\right)^{j}V({\bf R}).
\label{vm}
\end{eqnarray}
Equations (\ref{Dyson}) and (\ref{vm}) are exact in the limit $\omega_{c} \to
\infty$ and valid for any (differentiable) potential $V(X,Y)$. For the specific
case of a quadratic potential, only the first terms $k=0,1,2$ and $j=0,1$ of
the series appearing, respectively, in Eqs. (\ref{Dyson}) and (\ref{vm}) remain
giving rise to a nontrivial second-order partial differential equation to be
solved in Sec.~\ref{Section3}. Let us
first continue considering a generic potential and try to simplify at maximum
this Dyson equation.

In order to solve Eq. (\ref{Dyson}), it appears very convenient to introduce
modified vortex Green's function through the following change in functions
[an insight suggested by the form of the electronic Green's function (\ref{Gelec})]
\begin{eqnarray}
\tilde{g}_{m}^{R,A}({\bf R}) & = & e^{-\frac{l_{B}^{2}}{4} \Delta_{{\bf R}}} g^{R,A}_{m}({\bf R}) \label{change1} \\
&=&\sum_{p=0}^{+ \infty} \frac{1}{p!} \left(-\frac{l_{B}^{2}}{4} \Delta_{{\bf R}} \right)^{p}
g_{m}^{R,A}({\bf R}), \nonumber \\
\label{change2}
\tilde{v}_{m}({\bf R})&=& e^{-\frac{l_{B}^{2}}{4} \Delta_{{\bf R}}} v_{m}({\bf R}) \\
\nonumber
&=& \sum_{p=0}^{+\infty} a_{m,p} \left( \frac{l_{B}^{2}}{4}\Delta_{{\bf R}}\right)^{p} V({\bf R}),\\
a_{m,p}&=&
\sum_{j=0}^{p} \frac{(-1)^{p-j}}{(p-j)!} \frac{2^{j} (m+j)!}{m!(j!)^{2}},
\label{amp}
\end{eqnarray}
where expression~(\ref{vm}) for $v_{m}({\bf R})$ was used to obtain Eq. (\ref{amp}).
After some standard manipulations presented in Appendix~\ref{AppendixA}, one gets the
following very compact form of Dyson equation (valid for an arbitrary potential, in
the limit $\omega_c=\infty$ with $l_B$ finite):
\begin{eqnarray}
\left(
\omega -E_{m} \pm i \delta
\right)
\tilde{g}_{m}^{R,A}({\bf R})
=1 \hspace*{3cm}
 \nonumber \\ +e^{i \frac{l_{B}^{2}}{2}
\left(\partial_{X}^{\tilde{v}} \partial_{Y}^{\tilde{g}}- \partial_{Y}^{\tilde{v}} \partial_{X}^{\tilde{g}}\right)
}
\tilde{v}_{m}({\bf R}) \tilde{g}^{R,A}_{m}({\bf R})
,
\label{comp1}
\end{eqnarray}
where the notations $\partial_{X}^{\tilde{v}}$ and $\partial_{Y}^{\tilde{v}}$
mean that these spatial derivatives act on the function $\tilde{v}_{m}({\bf R})$ only
(similarly for $\tilde{g}_{m}({\bf R})$). Interestingly, and in contrast to the initial
Dyson Eq.~(\ref{Dyson}), this differential operator starts
now at order $l_B^4 \partial_{XY}\tilde{v} \partial_{XY}\tilde{g}$
(once Dyson equation has been properly symmetrized by taking its real part,
see Appendix~\ref{AppendixA}), so that the change in functions~(\ref{change1}) and (\ref{change2})
manages in principle to perform the whole resummation of potential gradient terms
to all orders in $l_B$ (this will be discussed in more detail in
Sec.~\ref{Section4}).

Before considering the solution of the transformed Dyson equation~(\ref{comp1}),
we need to examine the change brought by the mapping~(\ref{change1}) in the
electronic Green's function~(\ref{Gelec}), now diagonal in the Landau level
index $m$:
\begin{eqnarray}
G^{R,A}({\bf r},{\bf r}',\omega)
=
 \int \!\! \frac{d^{2} {\bf R}}{2 \pi l_{B}^{2}} \sum_{m=0}^{+
\infty}\tilde{g}_{m}^{R,A}({\bf R}) \nonumber \hspace*{1cm}
\\
\label{Greenexplicit2}
  \times
e^{ - \frac{l_{B}^{2}}{4}\Delta_{{\bf R}}} \left[
 \Psi_{m,{\bf R}}^{\ast}({\bf r}')\Psi_{m,{\bf R}}({\bf r}) \right],
\end{eqnarray}
where the factors $e^{-\frac{l_{B}^{2}}{2}\Delta_{{\bf R}}}$ and
$e^{\frac{l_{B}^{2}}{4}\Delta_{{\bf R}}}$ were combined together, and
integrations by parts were performed. The last step, performed in
Appendix \ref{AppendixB}, is simply to compute the action of the exponential
operator in Eq. (\ref{Greenexplicit2}) onto the product of two vortex wave
functions, which finally reads:
\begin{eqnarray}
e^{ - \frac{l_{B}^{2}}{4}\Delta_{{\bf R}}} \left[
 \Psi_{m,{\bf R}}^{\ast}({\bf r}')\Psi_{m,{\bf R}}({\bf r}) \right]
= \left| \Phi_{m}\left({\bf R}-[{\bf r}+{\bf r}']/2\right) \right|^{2} \nonumber \\ \times
e^{i \frac{{\bf R} \cdot [({\bf r}'- {\bf r})\times \hat{{\bf z}} ]}{l_{B}^{2}}}
e^{i \frac{yx'-xy'}{2 l_{B}^{2}}}, \label{Phi} \hspace*{0.5cm}
\end{eqnarray}
where
\begin{eqnarray}
\left|\Phi_{m}({\bf R}) \right|^{2} =
\frac{1}{\pi m! l_{B}^{2}} \frac{\partial^{m}}{\partial s^{m}}
\left.
\frac{e^{-A_{s}{\bf R}^{2}/l_{B}^{2}}}{1+s}
\right|_{s=0}
\label{trick}
\end{eqnarray}
with $A_{s}=(1-s)/(1+s)$. Form (\ref{Phi}) will be particularly useful for
subsequent calculations in Sec. \ref{Section4} and in Appendixes \ref{AppendixC}
and \ref{AppendixE}.

\section{Solving Dyson equation}
\label{Section3}

\subsection{Absence of curvature: case of an arbitrary 1D potential or a locally
flat disordered 2D potential}

Dyson equation (\ref{comp1}), also in its explicit form [Eq. (\ref{eqgen})], has the
remarkable property that the differential operators involve necessarily
derivatives of the potential in two orthogonal directions. For a 1D potential
along the $x$ direction, the function $v_{m}(X)$ depends on a single coordinate, so
that Dyson equation for $\tilde{g}_{m}^{R,A}(X)$
becomes completely trivial and its exact expression (in the limit
$\omega_{c} \to \infty$) reads
\begin{eqnarray}
 \tilde{g}_{m}^{R,A}(X)=\frac{1}{\omega-E_{m} -\tilde{v}_{m}(X) \pm i \delta},
\label{1Dexact}
\end{eqnarray}
with $\tilde{v}_{m}(X)$ defined above in Eq. (\ref{change2}) playing the role of
an effective potential energy.

To benchmark expression (\ref{1Dexact}) for the modified vortex Green's
function, we consider the exact solution for the electronic
Green's function that can be derived using a standard wave function formalism in
the case of a parabolic 1D potential and prove in Appendix \ref{AppendixC} that
both approaches lead to identical expressions. This establishes that
formula (\ref{Greenexplicit2}), with even the lowest order vortex Green's
function $\tilde{g}_m({\bf R})$, contains the edge states physics, which plays an
important role in the understanding of transport properties observed in the quantum
Hall effect regime.
\cite{Halperin1982,MacDonald1984,Buttiker1988}

From the present analysis of an arbitrary 1D potential, one can already guess
(see Sec.~\ref{Section4} for more details) that the differential operators
appearing in Dyson equation (\ref{comp1}) mainly play a role in the case of 2D
equipotential lines that present a certain amount of curvature at the scale of the
magnetic length. For a disordered 2D potential $V({\bf R})$, this can occur, e.g.,
in the vicinity of its critical points ${\bf R}_{c}$ characterized by
${\bm \nabla}V({\bf R}_{c})={\bf 0}$. For an arbitrary smooth potential, and far
from its critical points, the equipotential lines are locally straight at the scale
of $l_{B}$, so that the modified vortex Green's function $\tilde{g}_{m}^{R,A}({\bf R})$
will be well approximated by the expression
\begin{equation}
\tilde{g}_{m}^{R,A}({\bf R}) \approx \frac{1}{\omega - E_{m}-V({\bf R}) \pm i \delta} .
\label{approxh}
\end{equation}
Once inserted in the electronic Green's function (\ref{Greenexplicit2}), this
simple result gives the approximate expression
\begin{eqnarray}
G^{R,A}({\bf r},{\bf r}',\omega)
\approx
 \int \!\! \frac{d^{2} {\bf R}}{2 \pi l_{B}^{2}} \sum_{m=0}^{+ \infty}
\frac{ e^{ - \frac{l_{B}^{2}}{4}\Delta_{{\bf R}}} \left[
 \Psi_{m,{\bf R}}^{\ast}({\bf r}')\Psi_{m,{\bf R}}({\bf r}) \right]}{
\omega - E_{m}-V({\bf R}) \pm i \delta}
\nonumber \hspace*{-0.5cm} \\
\label{GTrugman}
\end{eqnarray}
that was proposed with little detail in our previous Ref. \onlinecite{Champel2008}.

Considering that for a smooth 2D potential the equipotential lines are locally straight
on the scale $l_{B}$ (this requires a sufficiently large local radius of curvature), one can then
perform in principle the integration over the variable parametrizing distance along the
constant energy ``surface'', $V({\bf R})=\mathrm{const}$, in the same way as done
explicitly in Appendix \ref{AppendixC} for a pure 1D potential. One then recovers from the
obtained Green's function expression the property that the wave functions are
locally well approximated by translation-invariant Landau states with drift velocity
$c {\bm \nabla}V \times \hat{{\bf z}}/(|e|B)$, as argued in the seminal paper by
Trugman. \cite{Trugman1983} Expression (\ref{GTrugman}) is however quite
powerful, because it does not rely on a particular parametrization of the
equipotential lines, which can be cumbersome for a disordered potential, and can be used
easily by numerically or analytically performing the integral over the vortex
coordinate ${\bf R}$.

However, as stressed before, approximation (\ref{GTrugman}) breaks down in the
vicinity of the critical points of the potential, where the drift velocity
locally vanishes. This requires to include in the analysis the second-order derivatives
of the potential $V$ in order to lift the degeneracy of the Landau levels,
leading to strong quantum effects (quantization and/or lifetime), as we will discuss
from now on.

\subsection{Green's functions including curvature effects: case of a 2D quadratic potential}

To investigate curvature effects and to determine more precisely under which
conditions approximation (\ref{approxh}) is valid, we expand the arbitrary
potential $V({\bf R})$ around a given point ${\bf R}_{0}$, up to quadratic order.
This expansion appears to be sufficient provided that the gradient and the
three possible second-order derivatives of the potential (locally) never vanish
simultaneously, a realistic assumption. We thus write
\begin{eqnarray}
V({\bf R})&=& V({\bf R}_{0})+\left({\bf R}-{\bf R}_{0}\right)\cdot {\bm \nabla}_{{\bf R}_{0}}V({\bf R}_{0}) \nonumber \\ &&+ \frac{1}{2}\left[\left({\bf R}-{\bf R}_{0}\right)\cdot {\bm \nabla}_{{\bf R}_{0}} \right]^{2}V({\bf R}_{0}).
\label{expan}
\end{eqnarray}
Inserting expression (\ref{expan}) into formula (\ref{change2}), we get
\begin{equation}
\tilde{v}_{m}({\bf R})=V({\bf R})+\frac{l_{B}^{2}}{2}\left( m+\frac{1}{2}\right) \left. \Delta_{{\bf R}} V \right|_{{\bf R}={\bf R}_{0}}.
\end{equation}
From the symmetrized form (\ref{eqgen}) of Dyson equation (\ref{comp1}), we then find that the
function $\tilde{g}_{m}({\bf R})$ is dictated by the second-order partial differential
equation
\begin{eqnarray}
\frac{l_{B}^{4}}{8}
\left[
 \left( \partial_{Y}^{2} V \right)\partial^{2}_{X}+\left(\partial_{X}^{2} V \right) \partial^{2}_{Y} - 2 \left(\partial_{X}\partial_{Y}V \right) \partial_{X} \partial_{Y}
\right]\tilde{g}_{m}^{R,A}({\bf R}) \nonumber \hspace*{-0.5cm} \\
+ \left[ \omega-E_{m} -\tilde{v}_{m}({\bf R}) \pm i \delta\right]\tilde{g}_{m}^{R,A}({\bf R})
=1. \hspace*{1cm} \label{eqquadra}
\end{eqnarray}
The antisymmetrized Dyson equation (\ref{eqgen2}) yields, on the other hand, the extra constraint
\begin{eqnarray}
\left( \partial_{X}V \right) \partial_{Y} \tilde{g}_{m}^{R,A} - \left( \partial_{Y}V  \right) \partial_{X} \tilde{g}_{m}^{R,A}=0
\end{eqnarray}
indicating that the function $\tilde{g}_{m}^{R,A}({\bf R})$ necessarily
possesses the same equipotential lines as $V({\bf R})$. We thus write
$\tilde{g}_{m}^{R,A}({\bf R})=f_{m}^{R,A}\left[E({\bf R})\right]$ where $E({\bf
R})=V({\bf R})-V({\bf R}_{0})$ and substitute this expression into Eq.
(\ref{eqquadra}) to obtain a simple 1D differential equation obeyed by the function
$f_{m}^{R,A}(E)$:
\begin{eqnarray}
\left[ \left(\gamma E +\eta\right)\frac{d^{2}}{dE^{2}}+ \gamma \frac{d}{dE}
\right]f_{m}^{R,A}(E) \nonumber \\ +\left[\tilde{\omega}_{m}-E \pm i \delta\right]f_{m}^{R,A}(E)=1,
\label{eqf}
\end{eqnarray}
with
\begin{eqnarray}
\tilde{\omega}_{m} &=&\omega-E_{m}-\tilde{v}_{m}({\bf R}_{0}),
\label{omegatilde}
\\
\gamma &= & \frac{l_{B}^{4}}{4}\left[\partial_{X}^{2}V \partial_{Y}^{2}V -\left( \partial_{X}\partial_{Y}V\right)^{2}
\right]_{{\bf R}={\bf R}_{0}}
,\\
\eta &=&\frac{l_{B}^{4}}{8}\left[\partial_{X}^{2}V \left(\partial_{Y}V\right)^{2}
+\partial_{Y}^{2}V \left(\partial_{X}V\right)^{2} \right. \nonumber \\
&& \left.
-2 \partial_{X}V\partial_{Y}V \partial_{X}\partial_{Y}V \frac{}{}\right]_{{\bf R}={\bf R}_{0}}
. \label{eta}
\end{eqnarray}
The coefficient $\gamma$ is nothing but the determinant of the Hessian matrix of the
potential $V$, with a prefactor $l_{B}^{4}/4$. Its sign determines the nature of the critical
points ${\bf R}_{c}$ at which $|\nabla_{\bf R} V|$ vanishes. A saddle point is
characterized by $\gamma({\bf R}_{c}) <0$, while a strictly positive $\gamma({\bf R}_{c})$
indicates the presence of a local maximum or minimum.

Differential equation~(\ref{eqf}) can be solved by Fourier transforming to time, see
Appendix~\ref{AppendixD}, so that the Green's function is given by the explicit
formula
\begin{eqnarray}
\tilde{g}_{m}^{R,A}({\bf R})=\int \!\! d t \, h_{m}^{R,A}({\bf R}_{0},t) \, e^{-i\left[V({\bf R})-V({\bf R}_{0}) \right] \tau(t)},
\label{resultath}
\end{eqnarray}
with
\begin{eqnarray}
\hspace{-0.6cm} h^{R,A}_{m}({\bf R}_{0},t)
& = &
\frac{ \mp i \theta\left(\pm t \right) }{\cos (\sqrt{\gamma} t )}
 e^{-i(\eta/\gamma) \tau(t)+
i(\tilde{\omega}_{m}+\eta/\gamma \pm i \delta)t}
\label{htilde}\\
\tau(t) & = & \frac{1}{\sqrt{\gamma}} \tan \left ( \sqrt{\gamma} t \right).
\label{tau}
\end{eqnarray}

Noticeably, when $\gamma>0$ the function $\tau(t)$ is a periodic function of
time $t$, so that Green's function $\tilde{g}_{m}$ must display discrete poles,
and quantization of energy levels in a confined potential is recovered (this is
further discussed in Appendix~\ref{AppendixE}). We stress that such success of
the vortex formalism was far from granted, because one has started with a family
of wave functions labeled by the continuous quantum number ${\bf R}$.

For $\gamma <0$, the functions $\cos$ and $\tan$ in Eqs. (\ref{htilde}) and
(\ref{tau}) are to be replaced by their hyperbolic counterparts $\cosh$ and
$\tanh$, respectively, so that the kernel $1/\cosh(\sqrt{-\gamma t})$
obviously introduces lifetime effects in the description [the convergence of 
integral (\ref{resultath}) over the time is now ensured by this term and no
more by the cutoff function $\exp(\mp \delta t)$, as was the case for
$\gamma > 0$]. Clearly, the vortex self-energy obtained from Eq.~(\ref{resultath}) displays
an elastic scattering rate proportional to $\sqrt{-\gamma}$, a clear signature
of quantum tunneling at saddle point with important consequences for transport
properties (see discussion in Sec.~\ref{Section5}). This allows us to make
the crucial physical identification between scattering mechanism and negative
curvature of the potential in the quantum Hall regime.

It is interesting to note that the strong quantum effects (quantization or lifetime) exhibited by the exact quantum solution (\ref{resultath}) are dictated by the quantitity $\sqrt{\gamma}$, which involves the square root of the second-order derivatives of the potential. Clearly, they thus can not be fully captured via a finite expansion in powers of the magnetic length  which can only generate integral powers of the derivatives of the potential; see Secs. \ref{section2b} and \ref{section2c}. This impossibility to approximate quantum effects at finite $l_{B}$ in a controllable way  with  the $l_{B}$ expansion illustrates its asymptotic character.

The function $h_{m}^{R,A}({\bf R}_{0},t)$ depends on the reference point ${\bf
R}_{0}$ via the coefficients $\eta$ and $\tilde{\omega}_{m}$ for a generic
quadratic potential, and possibly also via the coefficient $\gamma$ for a
potential characterized by higher derivatives. The geometric parameters
$\gamma$ and $\eta$ are basically small coefficients for a potential $V({\bf
R})$ which varies smoothly at the scale $l_{B}$. If we literally take
$\gamma=\eta=0$, we find again expression (\ref{approxh}) for the function
$\tilde{g}_{m}({\bf R})$. We will show further in which circumstance it is
nevertheless required to keep the dependence on the coefficients $\gamma$ and/or
$\eta$ in the Green's function to correctly describe the local physical
observables.

Making use of expression (\ref{resultath}) together with Eq.
(\ref{Greenexplicit2}), we obtain that the electronic Green's function reads
finally
\begin{eqnarray}
G^{R,A}({\bf r},{\bf r}',\omega)
=
 \int \!\! \frac{d^{2} {\bf R}}{2 \pi l_{B}^{2}} \sum_{m=0}^{+ \infty} e^{- \frac{l_{B}^{2}}{4}\Delta_{{\bf R}}} \left[
 \Psi_{m,{\bf R}}^{\ast}({\bf r}')\Psi_{m,{\bf R}}({\bf r})\right] \nonumber \hspace*{-0.5cm} \\
 \times
\int \!\! dt \, h_{m}^{R,A}({\bf R}_{0},t) \, e^{-i [V({\bf R})-V({\bf R}_{0})]\tau(t)}
\label{Greenexplicit}
. \hspace*{0.5cm}
\end{eqnarray}
Expression (\ref{Greenexplicit}) is the main mathematical result of this work.
It is exact in the limit $\omega_{c} \to \infty$ for any quadratic potentials.
In particular, it holds for quadratic confining potentials simulating closed
systems, such as quantum dots, as well as for nonconfining quadratic potentials
corresponding to open systems, such as quantum point contacts. Physical implications
of this result are discussed in Sec.~\ref{Section5}, while further
mathematical simplifications will now be performed in order to extract relevant
physical observables.

\section{Local density and curvature effects}
\label{Section4}

\subsection{Simplifying the Green's function expression}

Expression (\ref{Greenexplicit}) for the Green's functions can be calculated
further in different ways. One possibility is to use a parametrization of the
equipotential lines of $V({\bf R})$. Such an approach appears, however, not very
practical for a generic random potential. Actually, it turns out that the
two-dimensional integral over the position ${\bf R}$ can be performed
analytically when $V({\bf R})$ is expanded up to its second derivatives around
the point ${\bf R}_{0}$. For a quadratic potential, the Green's function can
be rewritten at the final stage as a single one-dimensional integral over the
time variable $t$, as will be shown in this section. For the numerics, this appears
to be more easily tractable than a direct computation of formula
(\ref{Greenexplicit}).

Note that for a quadratic potential, formula (\ref{Greenexplicit}) is actually
independent of the choice of ${\bf R}_{0}$. This can be easily checked by taking
the gradient of expression (\ref{Greenexplicit}) with respect to ${\bf R}_{0}$
and considering that, besides the explicit term $V({\bf R}_{0})$, the dependence
on ${\bf R}_{0}$ is also contained in the function $h_{m}({\bf R}_{0},t)$ through the
coefficients $\eta$ and $\tilde{\omega}_{m}$ (the other coefficient $\gamma$ is
independent of ${\bf R}_{0}$ in the particular case of a quadratic potential).
The independence of the electronic Green's function then follows from the
relation ${\bm \nabla}_{{\bf R}_{0}}\eta=\gamma \, {\bm \nabla}_{{\bf
R}_{0}}V({\bf R}_{0})$.

 For a smooth arbitrary potential $V({\bf R})$, result (\ref{Greenexplicit})
is expected to give a very good approximation to the electronic Green's function
provided that the temperature exceeds the energy scales associated with the third order (and beyond) derivatives of the potential. Contrary to the case of a quadratic potential, formula
(\ref{Greenexplicit}) will now depend on the reference point ${\bf R}_{0}$,
which thus has to be chosen appropriately. The natural choice appears to be
${\bf R}_{0}=({\bf r}+{\bf r}')/2$.

Inserting formulas (\ref{Phi}) and (\ref{trick}) into Eq. (\ref{Greenexplicit}),
using the expansion (\ref{expan}) of the potential $V({\bf R})$ up to quadratic
order with ${\bf R}_{0}=({\bf r}+{\bf r}')/2 ={\bf c}$ and evaluating the
resulting Gaussian integrals over the variable ${\bf R}$, we get
\begin{eqnarray}
G^{R,A}({\bf r},{\bf r}',\omega)
= \sum_{m=0}^{+ \infty} \int \!\! dt \,
\frac{h_{m}^{R,A}({\bf c},t)
}{2 \pi l_{B}^{2}} \frac{1}{m!} \, e^{i \frac{xy'-x'y}{2 l_{B}^{2}}}
\hspace*{0.5cm}
\nonumber \\ \times
 \frac{\partial^{m}}{\partial s^{m}}\left(
\frac{\exp
\left[
- \frac{\tau^{2}(t)}{4} \frac{A_{s} l_{B}^{2} \left|{\bf W}({\bf r},{\bf r}',t)\right|^{2}+4i \tilde{\eta}({\bf r},{\bf r}',t)\tau(t)}{A_{s}^{2}+i A_{s} \zeta \tau(t)-\gamma \tau^{2}(t)}
\right]}{(1+s)\sqrt{A_{s}^{2}+iA_{s}\zeta \tau(t)-\gamma \tau^{2}(t)}}
\right)_{s=0}
\label{Gresum}
\end{eqnarray}
with
\begin{eqnarray}
{\bf W}({\bf r},{\bf r}',t) &=& {\bm \nabla}_{{\bf c}} V({\bf c}) + \hat{{\bf z}} \times \frac{{\bf r}'-{\bf r} }{l_{B}^{2} \tau(t)} ,
\\
\tilde{\eta}({\bf r},{\bf r}',t) &=& \frac{l_{B}^{4}}{8} \left[ ({\bf W} \times \hat{{\bf z}})\cdot \left. H_{V} \right|_{{\bf c}} ({\bf W} \times \hat{{\bf z}}) \right],
\\
\zeta &=&\frac{l_{B}^{2}}{2} \Delta_{{\bf c}} V({\bf c}),
\end{eqnarray}
where $\left. H_{V} \right|_{{\bf c}}$ is the 2 x 2 Hessian matrix composed of
the second derivatives of the potential $V$ taken at position ${\bf c}$ (the
elements of the matrix are given by $[ H_{V}]_{ij}=\partial_{i}\partial_{j}V$).
For ${\bf r}={\bf r}'$, we have the simplifications ${\bf W}({\bf r},{\bf
r},t)={\bm \nabla}_{{\bf r}} V({\bf r})$ and $\tilde{\eta}({\bf r},{\bf
r},t)=\eta({\bf r})$ [function defined in Eq. (\ref{eta})]. Note that for a
potential $V$ characterized by derivatives of order higher than 2, formula
(\ref{Gresum}) yields only an approximate result. In this case, all the
geometric coefficients, including $\gamma$ and $\zeta$, depend on the center of
mass position ${\bf c}$.

\subsection{Local density of states}

We now aim at computing the local electronic density defined by
\begin{eqnarray}
n({\bf r})= \int \!\! d\omega \, n_{F}(\omega) \rho({\bf r},\omega),
\label{localdens}
\end{eqnarray}
where the local density of states $\rho({\bf r},\omega)$ is directly obtained
from the retarded Green's function at coincident positions as
\begin{eqnarray}
\rho({\bf r},\omega)= - \frac{1}{\pi} \mathrm{Im} [G^{R}({\bf r},{\bf r},\omega)].
\label{localdos}
\end{eqnarray}
Here $n_{F}(\omega)=[1+\exp([\omega-\mu]/T)]^{-1}$ is the Fermi-Dirac distribution function, $T$ the temperature,  and $\mu$ the chemical potential.
We thus need the simpler form of expression (\ref{Gresum})
\begin{eqnarray}
G^{R,A}({\bf r},{\bf r},\omega)
= \sum_{m=0}^{+ \infty} \int \!\! dt \,
\frac{h_{m}^{R,A}({\bf r},t)
}{2 \pi l_{B}^{2}} \frac{1}{m!} \hspace*{2cm} \nonumber \\ \times
 \frac{\partial^{m}}{\partial s^{m}}\left(
\frac{\exp
\left[
- \frac{\tau^{2}(t)}{4} \frac{A_{s} l_{B}^{2} |{\bm \nabla}_{{\bf r}} V({\bf r})|^{2}+4i\eta({\bf r})\tau(t)}{A_{s}^{2}+i A_{s} \zeta \tau(t)-\gamma\tau^{2}(t)}
\right]}{(1+s)\sqrt{A_{s}^{2}+iA_{s}\zeta \tau(t)-\gamma \tau^{2}(t)}}
\right)_{s=0}
\label{Gresumdiag}.
\end{eqnarray}

To simplify further the expression of the local density, it is then required to
consider the explicit expression (\ref{htilde}) for the function
$h_{m}^{R}({\bf r},t)$ and insert it into Eq. (\ref{Gresumdiag}). In order to do
the integral over $\omega$ in expression (\ref{localdens}), we first
introduce the change in variable $\omega'=\omega-\mu$ and decompose the
exponential factor in the numerator depending on $\omega'$ as $\exp(i
\omega't)=\cos(\omega't)+i \sin(\omega't)$. The integral over the energy
$\omega'$ in Eq. (\ref{localdens}) coming with the first term $\cos(\omega't)$
is then performed by writing the Fermi-Dirac distribution function as
\begin{eqnarray}
n_{F}(\omega'+\mu)=\frac{1}{2} \left[1-\tanh \left( \frac{\omega'}{2T}\right) \right].
\end{eqnarray}
On the other hand, the second contribution to the integral (\ref{localdens})
coming with the term $\sin(\omega't)$ is calculated by using the result
\begin{eqnarray}
\int_{- \infty}^{+\infty} \!\! \!\! \! d\omega' \, \frac{ \sin(\omega' t) }{1+e^{\omega'/T}}
= -\frac{ \pi T}{\sinh(\pi T t)}
.
\end{eqnarray}
Finally, we find that
the local density takes the form of a simple integral over the time $t$
\begin{widetext}
\begin{eqnarray}
\nonumber
n({\bf r})
&= &
\frac{1}{2 \pi l_{B}^{2}} \left[
\frac{1}{2}
+
\mathrm{Im}
\sum_{m=0}^{+\infty}
\int_{0}^{+\infty} \!\!\!\!\!\! dt \, \frac{T}{\sinh\left[\pi T t \right]} \frac{e^{it[\mu-E_{m}-(m+1/2)\zeta-V({\bf r})] } e^{i\frac{\eta({\bf r})}{\gamma}[t-\tau(t)]} }{\cos(\sqrt{\gamma}t)} \right.
\\
&&
\left.
\hspace*{5cm} \times
\frac{1}{m!}
 \frac{\partial^{m}}{\partial s^{m}}
\left(
\frac{\exp
\left[
- \frac{\tau^{2}(t)}{4} \frac{A_{s} l_{B}^{2} |{\bm \nabla}_{{\bf r}} V({\bf r})|^{2}+4i\eta({\bf r})\tau(t)}{A_{s}^{2}+i A_{s} \zeta \tau(t)-\gamma \tau^{2}(t)}
\right]}{(1+s)\sqrt{A_{s}^{2}+iA_{s}\zeta \tau(t)-\gamma \tau^{2}(t)}}
\right)_{s=0}
\right].
\label{finaldensity}
\end{eqnarray}
\end{widetext}
This formula is exact for any quadratic potential in the absence of Landau
level mixing. To illustrate this strong statement, we prove in Appendix
\ref{AppendixE} its equivalence with the expression for the local density 
that can be derived by standard means in the specific case of a
circular 2D parabolic confinement (note that we have already shown the
correspondence in the different case of a 1D parabolic potential at the level of
the Green's functions, see Appendix \ref{AppendixC}). This shows that quantization 
effects, i.e., the presence of a discrete energy spectrum, are 
fully captured in the vortex representation, despite not being completely 
explicit in formula~(\ref{finaldensity}). The latter equation
has thus a relatively general character since it contains under a compact and
unified form the cases of confining and nonconfining quadratic potentials.
Note that expression (\ref{finaldensity}) is naively problematic for the
saddle-point quadratic potential model because the energy spectrum in this case
is unbounded from below, but relative density variations are, on the other hand,
perfectly well defined.


Of particular interest is the derivative of the local density with respect to
the chemical potential which can be directly probed by the differential
tunneling conductance in a scanning tunneling spectroscopy (STS) experiment
(provided that the tip density of states is constant in the studied energy
range):
\begin{equation}
\rho^{STS}({\bf r},\mu,T)=\frac{\partial n({\bf r})}{\partial \mu}= \int d \omega \left[-n'_{F}(\omega) \right] \rho({\bf r},\omega)
.
\label{rhoSTS}
\end{equation}
At zero temperature, this yields the local density of states at the
chemical potential energy, $\rho({\bf r},\mu)$, since then
$-n'_{F}(\omega)=\delta(\omega-\mu)$.
Using formula (\ref{finaldensity}), we directly get
\begin{widetext}
\begin{eqnarray}
\rho^{STS}({\bf r},\mu, T)
&= &
\frac{1}{2 \pi l_{B}^{2}}
\mathrm{Re}
\sum_{m=0}^{+\infty}
\int_{0}^{+\infty} \!\!\!\!\!\! dt \, \frac{ T t}{\sinh\left[\pi T t \right]}
\frac{e^{it[\mu-E_{m}-(m+1/2)\zeta-V({\bf r})]} e^{i\frac{\eta({\bf r})}{\gamma}[t-\tau(t)]} }{\cos(\sqrt{\gamma}t)}
\nonumber
\\
&&
\hspace*{5cm} \times
\frac{1}{m!}
 \frac{\partial^{m}}{\partial s^{m}}
\left(
\frac{\exp
\left[
- \frac{\tau^{2}(t)}{4} \frac{A_{s} l_{B}^{2} |{\bm \nabla}_{{\bf r}}
V({\bf r})|^{2}+4i\eta({\bf r})\tau(t)}{A_{s}^{2}+i A_{s} \zeta \tau(t)-\gamma\tau^{2}(t)}
\right]}{(1+s)\sqrt{A_{s}^{2}+iA_{s}\zeta \tau(t)-\gamma \tau^{2}(t)}}
\right)_{s=0}.
\label{localdosfin}
\end{eqnarray}
\end{widetext}
Contrary to the local density formula, expression (\ref{localdosfin}) is
well defined for the saddle-point quadratic potential model because it involves
only states around the energy $\mu$. Formula (\ref{localdosfin}) for the local density of states is exact for any quadratic potential.  One may wonder about its accuracy for an arbitrary potential landscape which is smooth on the scale of the magnetic length.
 We shall investigate this question by
a careful quantitative analysis in the next subsection.

\subsection{Quantitative aspects: when do gradient and curvature corrections
need to be included?}

In order to illustrate on a concrete example how successive steps in the
resummation of leading derivatives of the potential really operate, we focus
here on the 2D  circular confining potential 
\begin{eqnarray}
V({\bf r}) = \frac{1}{2} m^\ast \omega_0^2 {\bf r}^2
\label{confcirc}
,
\end{eqnarray}
whose explicit solution is given by the so-called Fock-Darwin states (see
Appendix~\ref{AppendixE}), and investigate the temperature-dependent
local density of states~(\ref{rhoSTS}).

The simplest approximation scheme, which amounts to view the potential term (\ref{confcirc})
in a purely local manner, i.e., $V({\bf R}) \simeq V({\bf R}_{0})$, is obtained by
setting $|{\bm \nabla} V|=\zeta=\gamma=\eta=0$ in Eq. (\ref{localdosfin}).
This obviously recovers the usual semiclassical guiding center result:
\begin{eqnarray}
\nonumber
\rho^{STS}_\mathrm{sc}({\bf r},\mu, T)
&= &
\sum_{m=0}^{+\infty}
\int_{0}^{+\infty} \!\!\!\!\!\!
\frac{dt}{2 \pi l_{B}^{2}}
\, \frac{ T t \cos(t[\mu-E_{m}-V({\bf r})])}
{\sinh\left[\pi T t \right]}
\\
& = &
-\frac{1}{2 \pi l_{B}^{2}}
\sum_{m=0}^{+\infty}
n_F'[E_m+V({\bf r})].
\label{rhoSC}
\end{eqnarray}
This result is in fact accurate as long as one considers temperatures higher
than the energy scale associated to the drift motion, namely, $l_B
|{\bm \nabla}_{\bf R}V|$.  
At lower temperatures, the resummation of all leading
gradient contributions needs to be performed, which corresponds to considering
the potential as locally flat (in the geometrical sense):
$V({\bf R}) \simeq V({\bf R}_{0})+\left({\bf R}-{\bf R}_{0}\right)\cdot
{\bm \nabla}_{{\bf R}_{0}}V({\bf R}_{0})$.
This calculation can in fact be achieved with the previously obtained results,
setting $\zeta=\gamma=\eta=0$ in Eq. (\ref{localdosfin}):
\begin{eqnarray}
\nonumber
\rho^{STS}_\mathrm{grad}({\bf r},\mu, T)
&= &
\sum_{m=0}^{+\infty}
\int_{0}^{+\infty} \!\!\!\!\!\!
\frac{dt}{2 \pi l_{B}^{2}}
\, \frac{ T t \cos(t[\mu-E_{m}-V({\bf r})])}
{\sinh\left[\pi T t \right]} \\
&& \hspace{-0.4cm}\times
\frac{1}{m!}
 \frac{\partial^{m}}{\partial s^{m}}
\left(
\frac{\exp
\left[
- \frac{t^2}{4} \frac{l_{B}^{2} |{\bm \nabla}_{{\bf r}} V({\bf r})|^{2}}{A_{s}}
\right]}{(1+s)A_{s}}
\right)_{s=0} .
\label{rhoGRAD}
\end{eqnarray}
Clearly the scale $l_B|{\bm \nabla}_{\bf R}V|$ provides a cutoff in the above
integral, so that the single pole divergence associated to the derivative
of the Fermi-Dirac distribution function in Eq. (\ref{rhoSC}) is regularized.

\begin{figure}[h!]
\includegraphics[scale=0.99]{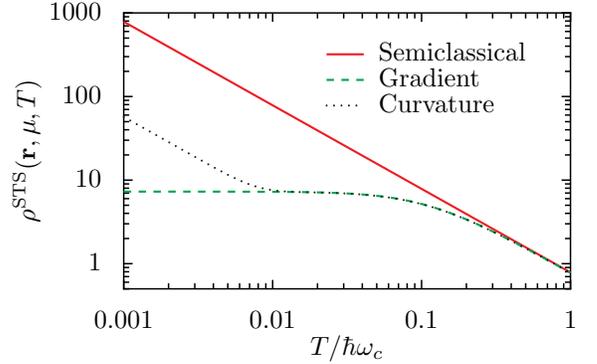}
\caption{(Color online) STS density of states at peak position ${\bf r}_{\mathrm{peak}}$,
in units of $(2\pi l_{B}^{2})^{-1}$, as a function of $T/\hbar\omega_c$, for
circular potential (\ref{confcirc}) with $\omega_0/\omega_c=0.1$, and $\mu/\hbar\omega_c=0.8$
(lowest Landau level). Semiclassical result (\ref{rhoSC}), gradient resummation
(\ref{rhoGRAD}), and full quadratic solution (\ref{localdosfin}) including curvature effects are presented.}
\label{fig1}
\end{figure}

\begin{figure}
\includegraphics[scale=0.99]{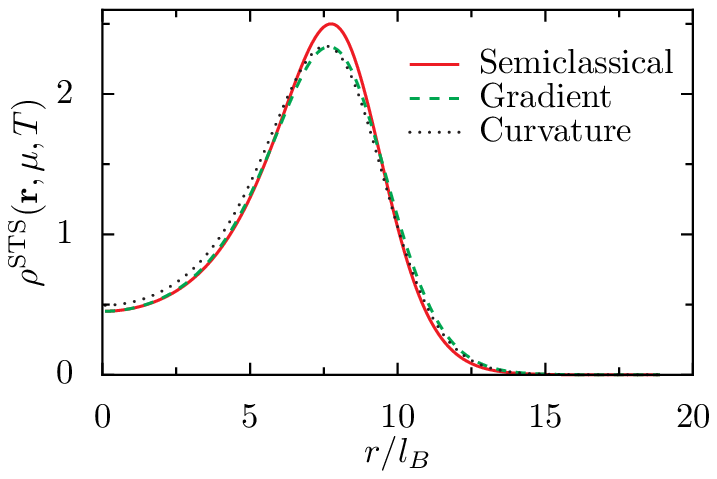}
\caption{(Color online) Similar data as in Fig. \ref{fig1}, now as a function
of radial distance $r/l_B$, for the high temperature $T/\hbar\omega_c=0.1$.}
\label{fig2}
\end{figure}

\begin{figure}
\includegraphics[scale=0.99]{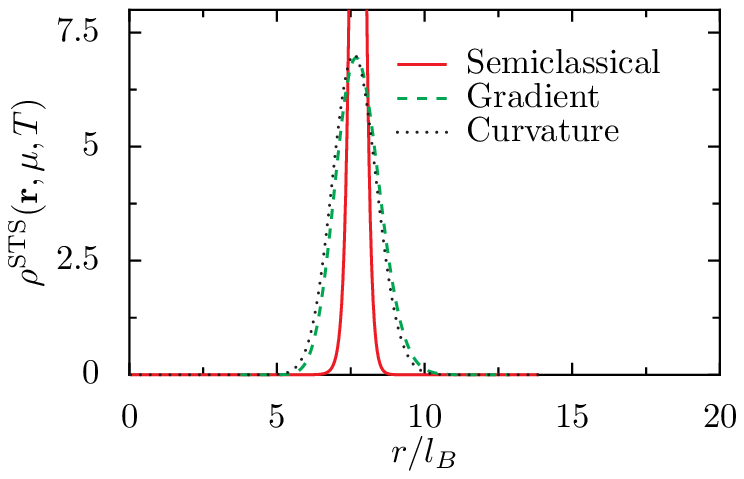}
\caption{(Color online) Similar data as in Fig. \ref{fig1}, now as a function
of radial distance $r/l_B$, for the intermediate temperature $T/\hbar\omega_c=0.01$.}
\label{fig3}
\end{figure}

\begin{figure}
\includegraphics[scale=0.99]{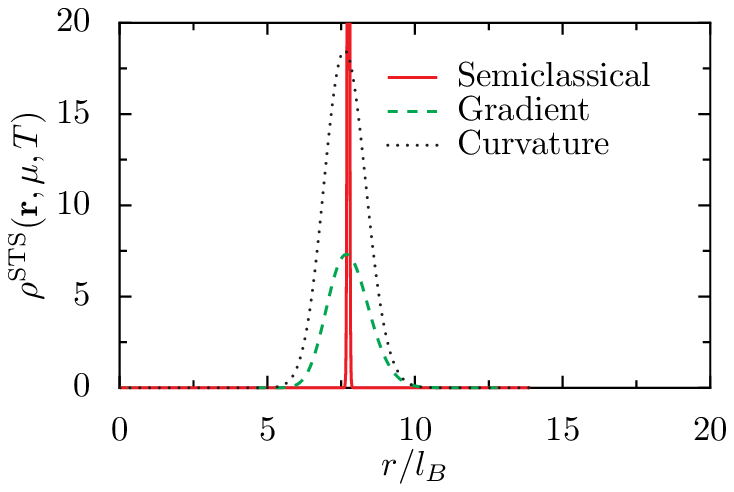}
\caption{(Color online) Similar data as in Fig. \ref{fig1}, now as a function
of radial distance $r/l_B$, for the low temperature $T/\hbar\omega_c=0.001$.}
\label{fig4}
\end{figure}

Fig. \ref{fig1} displays the STS local density of states as a function of temperature for a fixed chemical potential and the particular position ${\bf r}_{\mathrm{peak}}$ given by $\mu -E_{m}-V({\bf r}_{\mathrm{peak}})=0$, according to  semiclassical expression (\ref{rhoSC}), the leading gradient approximation (\ref{rhoGRAD}) and the exact solution (\ref{localdosfin}) which includes also curvature effects from the full quadratic dependence of potential (\ref{expan}). For the sake of simplicity, we have considered $\mu=0.8 \hbar \omega_{c}$, which corresponds to filling the lowest Landau level $m=0$ only.

Clearly in Fig. \ref{fig1}, the semiclassical approximation  is only valid in a high temperature regime and breaks down below the energy scale $l_B|{\bm \nabla}_{\bf R}V|$.
The departure of  semiclassical expression (\ref{rhoSC}) from the exact one [Eq. (\ref{localdosfin})] is also easily seen in the Figs. \ref{fig2} to \ref{fig4}, where the spatial dependence of the STS density of states is plotted for three different temperatures.
Expression (\ref{rhoGRAD}) which has a greater domain of validity than the semiclassical one turns out to match the exact result down to temperatures of the order of the curvature energy scale $\sqrt{\gamma}$.
The exponential cutoff in Eq. (\ref{rhoGRAD}) manifests itself 
 on Fig.~\ref{fig1} by a saturation at intermediate temperatures [where semiclassical result~(\ref{rhoSC}) is
already quite inaccurate]
of the STS density of states at peak value; see also Fig. \ref{fig3}. 
Decreasing further the temperature, curvature effects associated to the
small geometric energy scale $\sqrt{\gamma}$ begin to be felt; see Figs.~\ref{fig1} and \ref{fig4}. 
Departure from the leading gradient result is manifest by a final divergence of
the exact density of states at the peak position in the limit $T \to 0$, since the Fock-Darwin energy spectrum is discrete
(with level spacing $\sqrt{\gamma}$)  due to the confinement.

As a final remark, the above discussion is quite instructive, as it clearly shows under which conditions curvature effects associated with second-order derivatives of the potential can be neglected, namely, when temperature is higher than the energy associated with curvature.
Therefore,
successive approximation schemes  can be devised
for a smooth {\it arbitrary} (disordered or not) potential leading to controlled expressions for the local density of states. 
The whole scheme is indeed based on the existence of a hierarchy of local energy scales  of the type
$l_B^n \partial_{\bf r}^n V({\bf r})$. Expression (\ref{localdosfin}), which includes all second-order derivatives of the potential, thus provides an accurate estimate as long as the temperature is larger than cubic and higher order derivatives of the potential. In particular, it is also valid near saddle points of the potential landscape, where the drift velocity vanishes.
It is thus extremely useful for
interpreting local STS experiments such as in Ref. \onlinecite{Hashimoto2008} and completely
bypasses the need to diagonalize numerically a complicated random Schr\"odinger
equation.

\section{Discussion: on the fundamental importance of overcompleteness}
\label{Section5}

In the light of the technical results derived in the previous sections, we
formulate here some general conclusions on very fundamental issues in quantum
mechanics such as the emergence of classicality and the microscopic origin of
time irreversibility.

\subsection{Emergence of classicality in quantum mechanics}

It is well-known that the classical Hamilton-Jacobi equations of motion can be
derived from the quantum-mechanical Schr\"{o}dinger equation when terms having
$\hbar$ as prefactor can be disregarded. In other terms, classical mechanics is
clearly a limit of quantum mechanics. However, capturing the precise mechanism
responsible for the emergence of the classical behavior in the physical
properties of the system within a fully quantum mechanical framework, i.e., at
$\hbar$ finite, appears much more complicated. The essential reason is that
establishing the quantum-classical correspondence requires not only to consider
the equations of motion but also the states of the system. And the limit $\hbar
\to 0$ appears to be much more singular for the wave functions than for the
energy spectrum. When dealing with this limit, we are immediately confronted with
a conceptual problem relying on the fact that quantum mechanics is originally
formulated in a Hilbert space spanned by a countable basis of square integrable states,
while classical dynamics occurs in a continuous phase space. We are therefore in a
delicate position to reproduce the basic structure of the classical phase space.

In the particular problem under study in this paper, the set of vortex states
$|m,{\bf R} \rangle$ introduces from the very beginning in the quantum
description a continuous representation for the quantum numbers. Because they
obey in part the coherent states algebra \cite{Glauber} (note that the vortex
states are very peculiar coherent states in so far as they present the coherent
character only via the degeneracy quantum number ${\bf R}$ and not via the
eigenvalue quantum number $m$, so that they can be also eigenstates of the
kinetic part of the Hamiltonian, in contrast to fully coherent states), and
especially a completeness relation [Eq. (\ref{comp})], we can legitimately use
the vortex representation for the spectral decomposition of Hamiltonian
(\ref{Ham}) provided that the potential $V$ is a smooth function.
\cite{Champel2007} As an original motivation to work preferentially with these
states, \cite{Champel2007} the quantum numbers $|m,{\bf R} \rangle$ provide a
very intuitive and clear physical connection to the classical dynamics for the
free Hamiltonian when considering the de Broglie-Madelung hydrodynamic picture
\cite{Madelung,jap} of the Schr\"{o}dinger equation: the quantization of the kinetic energy into
Landau levels stems only from the interference of the electronic wave function
with itself due to the completion of a circular orbit around the position ${\bf
R}$, where the phase of the wave function is ill-defined. The price to pay for
the continuous aspect, i.e., for introducing overcompleteness into the quantum-mechanical formalism, is the nonorthogonality of the states with respect to
the degeneracy quantum number ${\bf R}$, which reflects the quantum
indeterminacy in the positions of the vortices and is accounted for in formula
($\ref{comp}$) by associating the elementary area $2 \pi l_{B}^{2}$ to the
incremental area in the integration over the vortex positions.

Being better armed to capture the transition from the quantum to classical, it
is not completely a surprise that we find that the vortex representation leads
at the mathematical level to a systematic and straightforward expansion
\cite{Champel2008} in powers of the magnetic length (which, we remind, plays the
role of an effective Planck's constant in the present problem) of the vortex
Green's functions, and thus of the physical observables. Therefore, it turns
out that overcompleteness is clearly not a drawback but an advantage at the
technical level! However, behind this mathematical aspect, we also see a very
fundamental physical aspect, which is rarely considered in quantum mechanics
when choosing a peculiar representation of states. Obviously, the vortex
representation offers the unique opportunity to derive quantum expressions
without having to implement the complete explicit form of the potential $V$.
This is exemplified by the exact compact formula (\ref{Greenexplicit}) for the
Green's function which embraces all possible cases of quadratic potentials. The
generic form of this result actually encodes the stability of the vortex states.
Indeed, the Fock-Darwin states (\ref{Fock}) which correspond to the exact
eigenstates of Hamiltonian (\ref{Ham}) in the presence of a circular parabolic
confinement and have the rotational symmetry (see Appendix \ref{AppendixE})
appear to be very unstable: one can not expect the confinement to be perfectly
circular under realistic conditions, so that the real physical state certainly
does not obey the rotational symmetry. In contrast, the vortex states which
enclose no preferred symmetry turn out to be stable with respect to an arbitrary
small asymmetrical smooth perturbation of the potential landscape. From this
robustness property, we can expect them to be the real physical states, i.e.,
the most predictable ones in an experiment.

Interestingly, we have an illustration with the present study for the process of
superselection of states put forward by Zurek \cite{Zurek2003} to explain the
emergence of the classical behavior from a quantum substrate. The only important
difference is that we are somewhat accounting here for an intrinsic mechanism of
classicality. Indeed, it is customary in quantum mechanics to appeal to
extrinsic degrees of freedom brought by an environment (surrounding the studied
quantum system) to explain the appearance of classical properties through
decoherence processes. As developed by several authors (see the review
\cite{Zurek2003}), the environment prevents certain quantum superposition of
states from being observed as a result of their high instability. Only states
that survive this process of coupling to the environmental degrees of freedom
have predictable consequences. As shown by Zurek {\em et al.} \cite{ZurekPRL} in
a model of weakly damped harmonic oscillator, coherent states, which are known
to be the closest states from the classical limit, are minimally affected by the
coupling to the environment. Due to this robustness, they emerge as a preferred
set of states.

In the present problem of the electron dynamics in a high magnetic field, 
we
clearly see under which conditions the overcomplete vortex representation becomes effectively
selected by the dynamics. 
Indeed, we have noted that formula
(\ref{Greenexplicit}) derived in the vortex representation reproduces the exact
Green's functions in the  simple integrable case  of a 2D circular confining potential (Appendix
\ref{AppendixE}). The system actually does not exhibit yet a preference for the
overcomplete set of vortex states over the complete set of Fock-Darwin eigenstates. 
In contrast, the case of a
quadratic saddle-point potential which simulates an open system and introduces a dynamical instability seems
quite instructive. Indeed, the conventional approach of quantum mechanics with
square integrable wave functions turns out to be inadequate to determine the energy spectrum, so that
one usually has to resort to another formalism, namely, the scattering states
quantum formalism. \cite{Fertig1987} These difficulties are manifestations of
the fact that the spectral problem for unstable unconfined dynamical systems
is not computable in the Hilbert space. The overcompleteness of the vortex representation in this
specific case of saddle-point potential shows precisely its relevance by
allowing one to solve the dynamical equations exactly on the same footing as in
the confining cases in a Green's function formalism. 
One can thus argue that the overcomplete set of vortex states is naturally 
favored by the instability of the dynamics.
Noticeably, the basic
dynamical object in the vortex representation appears to be no more the wave function but the Green's
function. By inspecting the form of the generic  Green's function (\ref{Greenexplicit}), one notices that the latter can not be written explicitly as a product of two wave functions (as is usually the case when using a complete representation) due to the presence of the nonlocal operator $\exp\left[-(l_{B}^{2}/4) \Delta_{{\bf R}}\right]$ acting on the vortex wave functions [see also Eqs. (\ref{Phi})-(\ref{trick})].
This reflects the overcompleteness of the coherent states basis with the two-dimensional continuous quantum numbers ${\bf R}$ associated to the vortex position.
It is therefore clear that it is not possible to get an single expression encompassing all possible cases of confining and unconfining quadratic potentials in terms of wave functions eigensolutions of the Schr\"{o}dinger's equation.  This general result can only be achieved through the introduction of an overcomplete basis of physical states.

\subsection{Time irreversibility}

An attractive feature is the close links existing between the transition from
quantum to classical (as a result of decoherence) and time irreversibility. By
time irreversibility we mean the time asymmetry due to a preferred direction of
time, as shown by decaying states. While quantum mechanics is able to provide a
clear and successful dynamical foundation to the idea of quantum levels, the
problem of decaying states with lifetimes remains somewhat obscure and
controversial. These difficulties in identifying the physical roots of
irreversibility rely essentially on the fact that the microscopic dynamical
equations are time reversible, whereas complex macroscopic systems are always
characterized by a time-asymmetric evolution. Consequently, it is generally
believed that irreversibility arises from the macroscopically large number of
degrees of freedom affecting the time evolution of a nonisolated system.
\cite{Lebowitz}

There have been many different approaches to derive an irreversible dynamical
evolution starting from the Schr\"{o}dinger equation. The most popular one
\cite{Braun} is to consider the microscopic (integrable) system as a part of a
larger Hamiltonian system which has many degrees of freedom (the environment or
reservoir). Then, after tracing over the environmental degrees of freedom (the
latter are disregarded because uncontrolled and unobserved), the dynamics of the
(open) quantum system is no more described by the Schr\"{o}dinger equation,
which is expected to be applicable only to a closed system. Other possibilities
are to solve quantum-mechanical equations by dealing directly with tractable
models of the environment, such as the consideration of a collection of harmonic
oscillators. The common denominator of all these approaches is to associate
time asymmetry with the external influence of a reservoir or a measurement
apparatus. Irreversibility thus seemingly has an extrinsic root.

In order to better clarify its possible link with the inherent dynamics of the
system, Prigogine {\em et al.} \cite{Prigogine1992,Prigogine1993,Prigogine1999}
demanded that irreversibility be rather directly connected with the Hamiltonian
of the microscopic quantum system, in spite of introducing extra dynamical
assumptions (because, after all, the division of a global system into a system and an environment is artificial and rather a matter of taste). These authors \cite{Prigogine1992,Prigogine1993,Prigogine1999} used
extensions of the traditional Hilbert space
through the introduction of a nonunitary change of representation and argued
with a few simple examples that time asymmetry may spontaneously arise in systems whose
dynamics is nonintegrable in the Hilbert space of quantum mechanics. Then, the
problem of integration and irreversibility both enjoy a common solution in the
extended space.

In the Hilbert space quantum mechanics, the time evolution described by the
Hamiltonian must be time reversible, leading to a widespread belief that
intrinsic irreversibility simply does not exist. Moreover, for the nontrivial
physically interesting systems, the computability of the spectral problem is
generally limited, the state of the art offering only perturbative and/or
effective approximate solutions. In such systems, irreversibility does appear in
the derivation, but as the result of supplementary approximations to the
Hamiltonian formalism of quantum mechanics. A well-known example in condensed-matter physics is the case of a disordered system for which elastic lifetimes in the
spectrum are obtained by averaging over disorder configurations.
\cite{Abrikosov} In brief, in order to clarify an intrinsic mechanism of
irreversibility, it is of valuable interest to find nontrivial physical systems
which are sufficiently simple to allow exact time-asymmetric solutions.

We strongly believe that the exact solution for the electron dynamics in a high
magnetic field and a given yet arbitrary quadratic potential presented in this paper
precisely offers such an opportunity. We have noted in Sec. \ref{Section3}
that the Green's functions are characterized by the presence of lifetimes in the
case of saddle-point potentials (when the geometric curvature $\gamma <0$),
meaning that time symmetry is broken. We thus obtained irreversibility without
appealing to extra dynamical considerations, such as an environmental coupling. In
other terms, we are basically in the scenario depicted by Prigogine {\em et al.}
\cite{Prigogine1992,Prigogine1993,Prigogine1999}

One may naturally wonder how the time-reversible Schr\"{o}dinger equation can
then lead to irreversible processes at the mathematical level. It is often
believed that the complex poles of the Green's functions correspond to
eigenvalues of a non-Hermitian operator. In contrast, we would like to point
out that a broken time symmetry exhibited by the states is not necessarily in
contradiction with a time-invariant Hamiltonian if a mathematical theory is
used that makes a distinction between states and the Hermitian Hamiltonian
operator. Actually, the dynamics remains here time symmetric but is realized
through an overcomplete representation which permits a broken time symmetry for
the states. A complete (countable) representation for its part does not allow
time-asymmetric solutions. The overcomplete vortex representation provides a
more general type of spectral decomposition of the Hamiltonian operator, which
is merely based on the use of Dirac's bra-ket formalism. The extension of the
eigenvalue problem to the complex plane is then purely a qualifying feature of
the instability of the dynamics, thus revealing an intrinsic irreversible
character of the evolution of the states.

It has been stressed by several authors
\cite{Prigogine1999,Bohm1998,Bohm1999,Delamadrid2005} that the natural setting of
quantum mechanics is the rigged Hilbert space rather than the Hilbert space
alone. The rigged Hilbert space is just an extended space consisting of the
Hilbert space equipped with distribution theory and was originally introduced
into quantum mechanics to give a mathematical justification of Dirac's bra-ket
formalism. It establishes rigorously that the spectral decomposition formula
acquires meaning in the continuous spectrum as well as in the discrete spectrum,
and allows the appearance of complex eigenvalues. Plane-wave eigenvectors,
which are generalized eigenvectors in the space of tempered distributions, are
basic examples of these elements of the rigged Hilbert space which do not live
in the Hilbert space. They are routinely used in the scattering states
formalism, which contains an arrow of time hidden in the choice of time
asymmetric boundary conditions: The consideration of in- and out- plane wave
states asymptotically far from the scattering region is indeed a statement of causality expressing the fact that the state at
a given position is determined by the action of a source at a retarded time.
Note that causality is naturally accounted for in the definition itself of the retarded and advanced Green's functions. However, in this case, the presence of the infinitesimal quantity $\delta$ in the dynamical equations [see Eq. (\ref{Greenevol})]  does not automatically imply a broken time symmetry for the physical states. For this, one needs in addition to have  a dynamical instability occurring in an unconfined system, i.e. scattering events.

The introduction of a continuous ingredient plays an important role in all
microscopic derivations of irreversible processes. With the consideration of
asymptotic in- and out-plane wave states, the scattering formalism presupposes
the existence of a continuum via the presence of reservoirs, so that
irreversibility finally acquires within this approach an extrinsic character.
Moreover, this formalism is specifically limited to open systems, thus
antagonistic to the Hilbert space quantum mechanics of closed systems. In this
paper, we have shown that, by using an overcomplete representation of coherent states
belonging to the Hilbert space such as the vortex states, it is possible to
embed quantum theory in a wider formalism of which Hilbert space quantum
mechanics of closed systems would become a special case. Moreover, in this approach quantization
effects and lifetime effects are naturally treated on the same footing. The
continuous ingredient is contained into the overcompleteness property of the
chosen set of quantum numbers. As a price to pay when working in a coherent states representation, it requires giving up the wave
functions as the fundamental quantity of quantum theory and replacing them by
Green's functions. It is worth emphasizing that the overcompleteness does not
necessarily imply a loss of information and time symmetry breaking. For this, we
need in addition an instability of dynamical motion related, e.g., to the presence of saddle points in the potential landscape. In this case, the
overcompleteness of the representation \cite{note} is necessary to obtain a solution of the
spectral problem. 
The basic reason is that the crossing
of the equipotential lines at the saddle-point energy (which schematically looks
like a collision process and can be seen as a bifurcation of a path) together with the openess of the system destroys the trajectory as well as the Hilbert space
description.
Therefore, the phenomenon of instability 
somehow imposes to deal directly with probabilities  to describe the dynamical evolution of the physical states (which necessarily belong to the Hilbert space).
It is worth noting that we then obtain a
description which from the point of view of its structure is isomorphic to
classical mechanics.

We have seen that irreversibility arises as a selection principle from the time-invariant Hamiltonian. The states selected by the unstable dynamics appear thus
to be less symmetric than they would seem to follow from the Hamiltonian description.
This situation is actually reminiscent of the well-known spontaneous symmetry
breaking as it occurs in ferromagnetism. In the presence of a dynamical
instability, bra and ket vortex states describe just physically distinct states.
Finally, we note that a critical ingredient to obtain the time symmetry breaking in our
solution is to consider quantum tunneling within an infinite system, i.e., unconfined
spatially (otherwise, the physical quantum numbers describing the dynamics are
necessarily discrete and the evolution unitary).

\section{Conclusion}

In this paper, we have built a Green's function formalism based on the use of an overcomplete semicoherent vortex representation to study the electron quantum dynamics in high magnetic fields and in a smooth potential landscape. Within this formalism, we have shown that it is possible to derive in a {\em controllable way} approximate quantum expressions, e.g., for the local density of states, for an arbitrary potential smooth at the scale of the magnetic length.
Moreover, we have obtained in the limit of negligible Landau level mixing an exact expression for the electronic Green's function which encompasses all possible cases of quadratic potentials. We have argued that this generic result, which is rendered possible by the use of an overcomplete representation of states belonging to the Hilbert space, is a manifestation of a stability property of the vortex quantum numbers.
We have shown that the overcompleteness feature of the vortex representation does not introduce de facto a loss of information, since we are able to reproduce the solutions for the exactly solvable (integrable) cases of parabolic  1D and 2D confining potentials, which can be obtained by standard wave function calculations.
In contrast, we have found that a loss of information, associated with the introduction of a probabilistic description of the physical processes, and concomitant with the appearance of lifetimes (synonymous of time symmetry breaking), arises in the saddle-point quadratic potential model. The vortex representation turns out to be especially relevant in this latter case of quadratic potential by providing in the limit of negligible Landau level mixing exact physical insight into the quantum tunneling processes originating at the saddle point.
Therefore, we have explicitly proved that time irreversibility does not result from supplementary
approximations to the Hamiltonian formalism of quantum mechanics, but just naturally 
arises in the spectral decomposition of the Hamiltonian from the formulation of dynamics in this overcomplete vortex representation of states.
With the present analysis, we deduce that the minimal necessary  ingredient to get solutions from the Hamiltonian formalism  which exhibit a broken time symmetry is to have an instability of the single-particle dynamics, as occurring from quantum tunneling at the saddle points of the potential landscape, which manifests itself in an unconfined (thus open) system.
Therefore, besides permitting to capture the transition from quantum to classical in an efficient way, the overcompleteness property of the representation allows the introduction of an intrinsic irreversibility on the
microscopic level.

\section*{Acknowledgement}
T.C. acknowledges interesting discussions with D. M. Basko.

\appendix

\section{Details on the mapping of Dyson equation in the high field limit}
\label{AppendixA}

Dyson equation~(\ref{Dyson}) has been rewritten in the $\omega_c=\infty$ limit
and we aim here at getting a simpler yet equivalent form that trivializes the
problem of local potential gradients. This can be achieved through the substitution
of functions~(\ref{change1}) and (\ref{change2}), which clearly gives
\begin{eqnarray}
\left( \omega -E_{m} \pm i \delta \right)
\tilde{g}_{m}^{R,A}({\bf R}) = 1+
\label{Dysontilde}
\hspace{3.3cm} \\
\sum_{k=0}^{+ \infty} \left( \frac{l_{B}}{\sqrt{2}} \right)^{2k}
\frac{1}{k!} e^{-\frac{l_{B}^{2}}{4} \Delta_{{\bf R}}}
\left[
\left( \partial_{X}-i\partial_{Y}\right)^{k}
e^{\frac{l_{B}^{2}}{4} \Delta_{{\bf R}}}
 \tilde{v}_{m}({\bf R}) \right.
\nonumber \\
\left.
\times
\left( \partial_{X}+i\partial_{Y}\right)^{k} e^{\frac{l_{B}^{2}}{4} \Delta_{{\bf R}}} \tilde{g}_{m}^{R,A}({\bf R})
\right]. \hspace*{0.5cm}
\nonumber
\end{eqnarray}
Going to Fourier space permits to rewrite the right-hand side of expression
(\ref{Dysontilde}) as a single global operator. Indeed, defining
\begin{eqnarray}
\tilde{g}_{m}({\bf R}) & = &\int \!\!\! d^{2} {\bf q} \, \tilde{g}_{m}({\bf q}) \, e^{i {\bf q} \cdot {\bf R}} ,\label{Fourier1}\\
\tilde{v}_{m}({\bf R}) & = &\int \!\!\! d^{2} {\bf p} \, \tilde{v}_{m}({\bf p}) \, e^{i {\bf p} \cdot {\bf R}}, \label{Fourier2}
\end{eqnarray}
and inserting these expressions into the right-hand side of Eq. (\ref{Dysontilde}),
important simplifications occur:
\begin{widetext}
\begin{eqnarray}
\left( \omega -E_{m} \pm i \delta \right) \tilde{g}_{m}^{R,A}({\bf R})-1= \nonumber\\
 \iint \!\! d^{2} {\bf q} \, d^{2} {\bf p}
 \sum_{k=0}^{+ \infty} \left( \frac{l_{B}}{\sqrt{2}} \right)^{2k}
\frac{1}{k!}
 \tilde{v}_{m}({\bf p}) \tilde{g}_{m}({\bf q}) \left[(ip_{x}+p_{y})(iq_{x}-q_{y}) \right]^{k} e^{\frac{l_{B}^{2}}{4} \left[ ({\bf p}+{\bf q})^{2}-{\bf p}^{2}-{\bf q}^{2}
\right]} e^{i ({\bf p}+{\bf q})\cdot {\bf R}} = \nonumber \\
 \iint \!\! d^{2} {\bf q} \, d^{2} {\bf p} \, \tilde{v}_{m}({\bf p}) \tilde{g}_{m}({\bf q}) \, e^{\frac{l_{B}^{2}}{2}(ip_{x}+p_{y})(iq_{x}-q_{y})} e^{\frac{l_{B}^{2}}{2} {\bf p} \cdot {\bf q}} e^{i ({\bf p}+{\bf q})\cdot {\bf R}} =
\nonumber
\\
 \iint \!\! d^{2} {\bf q} \, d^{2} {\bf p} \, \tilde{v}_{m}({\bf p}) \tilde{g}_{m}({\bf q}) \, e^{i\frac{l_{B}^{2}}{2}(p_{y}q_{x}-p_{x}q_{y})} e^{i ({\bf p}+{\bf q})\cdot {\bf R}}.
\end{eqnarray}
\end{widetext}
The global operator $e^{i l_{B}^{2}(p_{y}q_{x}-p_{x}q_{y})/2}$ above can
then be written back into real space, providing the final expression given
in Eq.~(\ref{comp1}).

We note in passing that the other Dyson equation ( i.e., $G=G_{0}+GVG_{0}$)
provides a second equation satisfied by the function $g_{m}$:
\begin{eqnarray}
\left(
\omega -E_{m} \pm i \delta
\right)
g_{m}^{R,A}({\bf R})
=
1+ \sum_{k=0}^{+ \infty} \left( \frac{l_{B}}{\sqrt{2}} \right)^{2k} \nonumber \\
\times
\frac{1}{k!}
\left( \partial_{X}+i\partial_{Y}\right)^{k} v_{m}({\bf R})
\left( \partial_{X}-i\partial_{Y}\right)^{k} g_{m}^{R,A}({\bf R})
\label{otherDyson},
\end{eqnarray}
which may be mapped in a similar way onto the following equation for the function $\tilde{g}_{m}$
\begin{eqnarray}
\left(
\omega -E_{m} \pm i \delta
\right)
\tilde{g}_{m}^{R,A}({\bf R})
=1\hspace*{3cm}
 \nonumber \\ +e^{-i \frac{l_{B}^{2}}{2}
\left(\partial_{X}^{\tilde{v}} \partial_{Y}^{\tilde{g}}- \partial_{Y}^{\tilde{v}} \partial_{X}^{\tilde{g}}\right)
}
\tilde{v}_{m}({\bf R}) \tilde{g}^{R,A}_{m}({\bf R}).
\label{comp2}
\end{eqnarray}
A more explicit expression for Dyson equation can then be obtained by taking the symmetric sum of
Eqs. (\ref{comp1}) and (\ref{comp2}), and afterward, by expanding the
exponential function and using the binomial theorem:
\begin{eqnarray}
\left(
\omega -E_{m} \pm i \delta
\right)
\tilde{g}_{m}^{R,A}({\bf R})
=1+\sum_{n=0}^{+ \infty} \left( \frac{l_{B}^{4}}{4}\right)^{n} \sum_{p=0}^{2n}
 \frac{(-1)^{p+n}}{p!(2n-p)!} \nonumber \hspace*{-0.5cm} \\ \times
 \partial_{X}^{p}\partial_{Y}^{2n-p} \tilde{v}_{m}({\bf R})
\partial_{X}^{2n-p}\partial_{Y}^{p} \tilde{g}_{m}^{R,A}({\bf R}). \hspace*{0.5cm}
\label{eqgen}
\end{eqnarray}
Note that the difference of Eqs. (\ref{comp1}) and (\ref{comp2}) yields another
equation which may be useful in solving Eq. (\ref{eqgen}) (e.g., in the case of
a quadratic potential, see Sec.~\ref{Section3})
\begin{eqnarray}
0=\sum_{n=0}^{+ \infty} \left( \frac{l_{B}^{4}}{4}\right)^{n} \sum_{p=0}^{2n+1}
 \frac{(-1)^{p+n}}{p!(2n+1-p)!} \nonumber \\ \times \partial_{X}^{p}\partial_{Y}^{2n+1-p} \tilde{v}_{m}({\bf R})
\partial_{X}^{2n+1-p}\partial_{Y}^{p} \tilde{g}_{m}^{R,A}({\bf R}).
\label{eqgen2}
\end{eqnarray}

\section{Modified vortex wave functions}
\label{AppendixB}

Our aim in this appendix is to
prove expression (\ref{Phi}). Let us analyze first the following differential operator:
\begin{eqnarray}
\hat{O}=\sum_{p=0}^{+\infty} \frac{1}{p!} \left( -\frac{l_{B}^{2}}{4} \Delta_{{\bf R}}\right)^{p}.
\end{eqnarray}
Applying this to a function $f({\bf R})$ and introducing the Fourier transform of $f$, we get
\begin{eqnarray}
\hat{O}[ f({\bf R})]
&=&\sum_{p=0}^{+\infty} \frac{1}{p!} \left( -\frac{l_{B}^{2}}{4} \Delta_{{\bf R}}\right)^{p} f({\bf R}) \\
&=& \int \!\!\! \frac{d^{2} {\bf q}}{(2 \pi)^{2}}
\sum_{p=0}^{+\infty} \frac{1}{p!} \left( \frac{l_{B}^{2}}{4} {\bf q}^{2} \right)^{p} \tilde{f}({\bf q}) \, e^{i {\bf q} \cdot {\bf R}}
\\
&=& \int \!\!\! \frac{d^{2} {\bf q}}{(2 \pi)^{2}}
 \tilde{f}({\bf q}) \, e^{i {\bf q} \cdot {\bf R}} \, e^{l_{B}^{2} {\bf q}^{2}/4}.
\end{eqnarray}
Using the inverse Fourier transform, we have
\begin{eqnarray}
\hat{O}[ f({\bf R})]
= \int \!\!\! d^{2} {\bf u} \, f({\bf u}) \int \!\!\! \frac{d^{2} {\bf q}}{(2 \pi)^{2}}
 \, e^{i {\bf q} \cdot \left( {\bf R} -{\bf u} \right)} \, e^{l_{B}^{2} {\bf q}^{2}/4}.
\end{eqnarray}
The integral over ${\bf q}$ is formally divergent. We circumvent this problem by
introducing for a while the parameter $\xi=-l_{B}^{2}/4>0$. The calculation of
the resulting Gaussian integral can then be easily done, which finally yields
\begin{eqnarray}
\hat{O}[ f({\bf R})]
= -\int \!\!\! \frac{d^{2} {\bf u}}{\pi l_{B}^{2}} \, f({\bf u}) \, e^{({\bf R}-{\bf u})^{2}/l_{B}^{2}}.
\label{convolu}
\end{eqnarray}
We deduce from this calculation that the operator $\hat{O}$ is nothing but a
convolution operator with a Gaussian kernel. We now apply it to $f({\bf R})=
\Psi_{m,{\bf R}}^{\ast}({\bf r}') \Psi_{m,{\bf R}}({\bf r})$. Using formula
(\ref{convolu}) and the explicit expression (\ref{vortex}) of the vortex wave
functions, we have
\begin{eqnarray}
\hat{O}[ \Psi_{m,{\bf R}}^{\ast}({\bf r}') \Psi_{m,{\bf R}}({\bf r}) ]
= -\frac{ e^{-({\bf R}-{\bf c})^{2}/l_{B}^{2}}
e^{i \left[(2 {\bf R}-{\bf c})\times {\bf d} \right]\cdot \hat{{\bf z}}/l_{B}^{2}}}{2 \pi l_{B}^{2}m!}
\nonumber \\
\times
\int \!\!\! \frac{d^{2} {\bm \eta}}{\pi l_{B}^{2}} \, \left[\frac{{\bm \eta}^{2}}{2 l_{B}^{2}} \right]^{m} \, e^{({\bm \eta}-2[{\bf R}-{\bf c}])^{2}/2l_{B}^{2}},\hspace*{0.5cm}
\label{conv}
\end{eqnarray}
where we have done the change in variable ${\bm \eta}={\bf u}-{\bf c}+i{\bf d}
\times \hat{{\bf z}}$ with ${\bf d}=({\bf r}'-{\bf r})/2$ and ${\bf c}=({\bf
r}'+{\bf r})/2$. We are again in presence of a formally divergent integral. As
just above we introduce the parameter $\xi$ and use the following trick to
perform the Gaussian integral over ${\bm \eta}$ in Eq. (\ref{conv}):
\begin{eqnarray}
-\int \!\!\! \frac{d^{2} {\bm \eta}}{\pi l_{B}^{2}} \, \left[\frac{{\bm \eta}^{2}}{2 l_{B}^{2}} \right]^{m} \, e^{({\bm \eta}-2[{\bf R}-{\bf c}])^{2}/2l_{B}^{2}}
= \nonumber \\
\int \!\!\! \frac{d^{2} {\bm \eta}}{4 \pi \xi} \, \left[\frac{-{\bm \eta}^{2}}{8 \xi } \right]^{m} \, e^{-({\bm \eta}-2[{\bf R}-{\bf c}])^{2}/8 \xi} = \nonumber \\
 \frac{\partial^{m}}{\partial s^{m}} \left.
\int \!\!\! \frac{d^{2} {\bm \eta}}{4 \pi \xi} \, e^{-s {\bm \eta}^{2}/8\xi} \, e^{-({\bm \eta}-2[{\bf R}-{\bf c}])^{2}/8 \xi}
\right|_{s=0}.
\label{lastint}
\end{eqnarray}
The remaining Gaussian integral (\ref{lastint}) can now be straightforwardly
evaluated (note that the contours of integration can be deformed to the real
axes using the analyticity property of the integrand). We finally find
\begin{eqnarray}
\hat{O}[ \Psi_{m,{\bf R}}^{\ast}({\bf r}') \Psi_{m,{\bf R}}({\bf r}) ]
=
\frac{ e^{-({\bf R}-{\bf c})^{2}/l_{B}^{2}}
e^{i \left[(2 {\bf R}-{\bf c})\times {\bf d} \right]\cdot \hat{{\bf z}}/l_{B}^{2}}}{\pi l_{B}^{2}m!} \nonumber \\
\times
\frac{\partial^{m}}{\partial s^{m}}
\left(
\frac{e^{\frac{2s}{1+s} \frac{({\bf R}-{\bf c})^{2}}{l_{B}^{2}}}}{1+s}
\right)_{s=0}
\label{convfin}. \hspace*{0.5cm}
\end{eqnarray}
Inserting the definitions of the parameters ${\bf c}$ and ${\bf d}$ in terms of
the positions ${\bf r}$ and ${\bf r}'$ into Eq. (\ref{convfin}), we directly
arrive at expressions (\ref{Phi}) and (\ref{trick}).

\section{Checking the vortex formalism: case of a 1D parabolic confining potential}
\label{AppendixC}

\subsection{Standard derivation}
In the particular case of a 1D parabolic potential given by
\begin{equation}
V(x)=\frac{1}{2}m^{\ast} \omega_{0}^{2}x^{2}
,
\label{parabolic1D}
\end{equation}
 the wave functions and the energy spectrum of Hamiltonian (\ref{Ham}) can be
found by solving directly the Schr\"{o}dinger equation using well-known standard
methods. The relevant quantum numbers appear to be a positive integer $n$ which
labels the Landau levels, and a continuous quantum number $p_{y}$ playing the role
of momentum in the $y$ direction. In the Landau gauge ${\bf A}=Bx \hat{{\bf y}}$,
wave functions and energy spectrum read, respectively,
\begin{eqnarray}
\Psi_{np_{y}}({\bf r}) &=&\frac{e^{-[x-(\omega_{c}/\Omega)p_{y}L^{2}]^{2}/2L^{2}}}{\sqrt{2^{n+1}n! \, \pi^{3/2} L}} H_{n}\left(\frac{x-\frac{\omega_{c}}{\Omega}p_{y}L^{2} }{L} \right) \nonumber \\
&& \times
e^{-ipy}
,
\label{wave1D}\\
E_{np_{y}} & = & \hbar \Omega \left(n+\frac{1}{2} \right) +V(p_{y}L^{2}),
\label{E1D}
\end{eqnarray}
where $\Omega=\sqrt{\omega_{c}^{2}+\omega_{0}^{2}}$ and
$L=\sqrt{\hbar/m^{\ast}\Omega}$ are the renormalized cyclotron pulsation and
magnetic length, respectively, and $H_{n}$ denotes the $n$th Hermite polynomial.

In absence of Landau level mixing, one has to consider $\omega_c\gg\omega_0$,
keeping all terms of order $\omega_0/\omega_c$, and neglecting higher powers of
this ratio. Thus we have $\Omega \approx \omega_{c}+\omega_{0}^{2}/2 \omega_{c}$
and $L \approx l_{B}$. From Eqs. (\ref{wave1D}) and (\ref{E1D}), the Green's function
thus reads in this limit of negligible Landau level mixing
\begin{eqnarray}
G^{R,A}({\bf r},{\bf r}',\omega)= \sum_{n=0}^{+\infty} \int \!\! d p_{y} \, \frac{e^{i p_{y}(y'-y)}}{\omega-E_{np_{y}}\pm i \delta }
\nonumber \\ \times
\frac{ e^{-\left[(x-p_{y}l_{B}^{2})^{2}+(x'-p_{y}l_{B}^{2})^{2}\right] /2l_{B}^{2}} }{2^{n+1} n! \pi^{3/2} l_{B}} \nonumber \\
\times H_{n}\left(\frac{x-p_{y}l_{B}^{2}}{l_{B}} \right) H_{n}\left(\frac{x'-p_{y}l_{B}^{2}}{l_{B}} \right)
\label{exactpara}, \hspace*{0.5cm}
\end{eqnarray}
with
\begin{equation}
E_{np} \approx \hbar \omega_{c} \left(n+\frac{1}{2} \right) + \hbar \omega_{0} \frac{\omega_{0}}{2 \omega_{c}} \left( n+\frac{1}{2}\right)
+V(p_{y}l_{B}^{2}).
\end{equation}

\subsection{Derivation within the vortex formalism}

Now, we show how one can recover the Green's function (\ref{exactpara}) of a 1D
parabolic potential from the vortex formalism. We start with expressions
(\ref{Greenexplicit2}) and (\ref{1Dexact}) and exploit the fact that the
effective potential $\tilde{v}_{m}$ is independent of the variable $Y$
\begin{eqnarray}
G^{R,A}({\bf r},{\bf r}',\omega)= \sum_{m=0}^{+\infty}
\int \!\! \frac{dX }{2 \pi l_{B}^{2}} \, \frac{1}{\omega- E_{m}-\tilde{v}_{m}(X) \pm i \delta} \nonumber \\
\times \int \!\! dY \, e^{ - \frac{l_{B}^{2}}{4}\Delta_{{\bf R}}} \left[
 \Psi_{m,{\bf R}}^{\ast}({\bf r}')\Psi_{m,{\bf R}}({\bf r}) \right].
\label{st1} \hspace*{0.5cm}
\end{eqnarray}
The integral over $Y$ can then be performed exactly making use of expressions (\ref{Phi}) and (\ref{trick}).
Considering that
\begin{eqnarray}
\int \!\! dY
e^{-A_{s} \frac{\left(Y- c_{y} \right)^{2}}{l_{B}^{2}}
-i \frac{(2Y-c_{y})d_{x}}{l_{B}^{2}}}
= \sqrt{\frac{\pi l_B^2}{A_{s}}} e^{-\frac{d_{x}^{2}}{A_s l_{B}^{2}}
-i \frac{c_{y}d_{x}}{l_{B}^{2}}}
\end{eqnarray}
where ${\bf c}$ and ${\bf d}$ are defined in Appendix \ref{AppendixB},
Eq. (\ref{st1}) is rewritten as
\begin{eqnarray}
G^{R,A}({\bf r},{\bf r}',\omega)= \sum_{m=0}^{+\infty}
\int \!\! \frac{ dX}{2 \pi l_{B}^{2}} \, \frac{1}{\omega- E_{m}-\tilde{v}_{m}(X) \pm i \delta} \nonumber \\
\times \frac{e^{2iXd_{y} /l_{B}^{2}} }{\sqrt{\pi} l_{B} m!} \,
e^{i \frac{xy-x'y'}{2 l_{B}^{2}}} \nonumber \\
\times
\frac{\partial^{m}}{ \partial s^{m}} \left[ \frac{1}{\sqrt{1-s^{2}}}
e^{-\frac{1}{A_{s}} \frac{d_{x}^{2}}{l_{B}^{2}}} e^{-A_{s} \frac{(X-c_{x})^{2}}{l_{B}^{2}}}
\right]_{s=0}
. \hspace*{0.3cm}
\label{st2}
\end{eqnarray}
It can be checked that the following algebraic relation holds:
\begin{eqnarray}
\frac{\partial^{m}}{\partial s^{m}}
 \left[
\frac{1}{\sqrt{1-s^{2}}}
e^{- \frac{1}{A_{s}} \frac{d_{x}^{2}}{l_{B}^{2}}} \,
e^{-A_{s} \frac{(X-c_{x})^{2}}{l_{B}^{2}}}
\right]_{s=0}
= \nonumber \\
2^{-m}
e^{-\frac{(X-c_{x})^{2}}{l_{B}^{2}}} e^{-\frac{d_{x}^{2}}{l_{B}^{2}}}
H_{m}\left(\frac{x-X}{l_{B}}\right) \, H_{m}\left(\frac{x'-X}{l_{B}} \right).
\label{nontrivial}
\end{eqnarray}
Inserting formula (\ref{nontrivial}) into Eq. (\ref{st2}) and reintroducing the
variables ${\bf r}$ and ${\bf r}'$ everywhere in place of ${\bf c}$ and ${\bf
d}$, we find that the Green's function finally reads

\begin{eqnarray}
G^{R,A}({\bf r},{\bf r}',\omega)= \sum_{m=0}^{+\infty}
\int \!\! \frac{ dX}{l_{B}^{2}} \, \frac{1}{\omega- E_{m}-\tilde{v}_{m}(X) \pm i \delta } \nonumber \\
\times e^{i \frac{X(y'-y)}{l_{B}^{2}}} \, \frac{ e^{-\left[(x-pl_{B}^{2})^{2}+(x'-pl_{B}^{2})^{2}\right] /2l_{B}^{2}} }{2^{m+1} m! \pi^{3/2} l_{B}}
\nonumber \\
 \times H_{m}\left(\frac{x-X}{l_{B}}\right) \, H_{m}\left(\frac{x'-X}{l_{B}} \right)
\, e^{i \frac{xy-x'y'}{2 l_{B}^{2}}}
. \hspace*{0.5cm}
\label{st3}
\end{eqnarray}
Introducing $p_{y}=X/l_{B}^{2}$ and expliciting the term $\tilde{v}_{m}(X)$ by
inserting expression (\ref{parabolic1D}) into definition (\ref{change2}) of
$\tilde{v}_{m}$, we see that expression (\ref{st3}) corresponds exactly to Eq.
(\ref{exactpara}) up to a phase factor $\exp\left[i (xy-x'y')/2 l_{B}^{2}
\right]$ which comes from the fact that we work here within the vortex formalism
in the symmetric gauge, and not in the Landau gauge.

\section{Solving the dynamical equation for potential lines}
\label{AppendixD}

Differential Eq. (\ref{eqf}) is second order in the derivative with respect
to $E$, but first order in $E$. It will obviously become second order in $\tau$
and first order in the derivative with respect to $\tau$ by going to the Fourier
component $F_{m}^{R,A}(\tau)$. So, in order to solve Eq. (\ref{eqf}), we write
\begin{equation}
 f_{m}^{R,A}(E)=\int \!\! d\tau \,
F_{m}^{R,A}(\tau) \, e^{-i E \tau},
\label{Fourier}
\end{equation}
and substitute this form into Eq. (\ref{eqf}) to get
\begin{eqnarray}
\int \!\! d\tau \, F_{m}^{R,A}(\tau)
\left[
\tilde{\omega}_{m} -E \pm i \delta-i \gamma \tau- (\gamma E+\eta)\tau^{2}
\right]
\nonumber e^{-i E \tau}
\\
\hspace*{-0.2cm} =\int \!\! d\tau \, F_{m}^{R,A}(\tau)
\left[
\tilde{\omega}_{m} \pm i \delta-i \gamma \tau - \eta \tau^{2} -i (\gamma \tau^{2}+1)\frac{d}{d\tau}
\right] \nonumber \hspace*{-0.2cm} \\
\times
e^{-i E \tau}
=1. \hspace*{6cm}
\end{eqnarray}
Doing an integration by parts,
we have
\begin{eqnarray}
1= \left[ -i(1+\gamma \tau^{2}) F_{m}^{R,A}(\tau) \, e^{-iE \tau}\right]_{- \infty}^{+\infty}
+\int \!\! d\tau \, e^{-i E \tau} \nonumber \\
\times
\left[
\tilde{\omega}_{m} \pm i \delta+i \gamma \tau- \eta \tau^{2} + i \left( 1+\gamma \tau^{2} \right) \frac{d}{d \tau}
\right] F_{m}^{R,A}(\tau)
. \hspace*{0.3cm}
\label{integrated}
\end{eqnarray}
Finally, taking the Fourier transform of this equation, we find that
$F_{m}^{R,A}(\tau)$ is governed by the first-order differential equation
\begin{eqnarray}
i \left[1+\gamma \tau^{2} \right] \frac{dF^{R,A}_{m}(\tau)}{d \tau} \hspace*{4cm} \nonumber \\
+\left[ \tilde{\omega}_{m} - \eta \tau^{2}+i \gamma \tau \pm i \delta \right]F^{R,A}_{m}(\tau)=
\delta ( \tau),
\label{nice}
\end{eqnarray}
provided that the integrated term in Eq. (\ref{integrated}) vanishes, i.e.,
\begin{equation}
(1+\gamma \tau^{2})F_{m}^{R,A}(\tau) \to 0
\label{condition}
\end{equation}
 when $\tau \to \pm \infty$.
Equation (\ref{nice}) is readily solved by
\begin{equation}
F_{m}^{R,A}(\tau)= \frac{\mp i\theta( \pm \tau )}{\sqrt{1+\gamma \tau^{2}}} \exp\left[i \left(\tilde{\omega}_{m}
+\eta/\gamma \pm i \delta\right) t(\tau) - i \frac{\eta}{\gamma} \tau
\right], \label{Fm}
\end{equation}
where we have introduced
\begin{eqnarray}
t(\tau) &=&\frac{1}{\sqrt{\gamma}} \arctan \left(\sqrt{\gamma} \tau \right).
\end{eqnarray}
Here $\theta(\tau)$ is the Heaviside function.
For $\gamma <0$, one must understand that
\begin{eqnarray}
\frac{1}{\sqrt{\gamma}} \arctan \left(\sqrt{\gamma} \tau \right)
&=&
\frac{1}{\sqrt{-\gamma}} \operatorname{artanh} \left(\sqrt{-\gamma} \tau \right) \nonumber \\
&=&
\frac{1}{2\sqrt{-\gamma}} \ln \left( \frac{1+\sqrt{- \gamma}\tau}{1-\sqrt{- \gamma} \tau} \right),
\end{eqnarray}
defined for $\sqrt{-\gamma} |\tau | \leq 1$. The variable $t$ in expression
(\ref{Fm}) plays actually the role of the time since it is conjugated to the
energy $\omega$ which enters into the expression via the quantity
$\tilde{\omega}_{m}$; see definition (\ref{omegatilde}). Because the solutions
of the homogeneous equation do not respect the time causality, one has only
considered the particular solution of inhomogeneous equation (\ref{nice}).

For $\gamma \leq 0$, solution (\ref{Fm}) fulfils requirement
(\ref{condition}) for any value of the parameter $\eta$ (for the case
$\gamma=0$, condition (\ref{condition}) is obeyed with the help of the
infinitesimal quantity $\pm i \delta$, while for $\gamma <0$ we have
$F_{m}^{R,A}=0$ for $\sqrt{-\gamma}| \tau| \geq 1$). However, for $\gamma > 0$,
we note that condition (\ref{condition}) is not satisfied, so that
expression (\ref{Fourier}) together with formula (\ref{Fm}) does not yield a
solution of the initial Eq. (\ref{eqf}). Nevertheless, the solution of Eq.
(\ref{eqf}) for $\gamma > 0$ can be inferred from result (\ref{Fm}) by
noting that the problem actually originates from the saturation of the function
$t(\tau)$ when $\tau \to \pm \infty$. Indeed, by considering $t$ instead of
$\tau$ as being the relevant variable and by extending its domain of definition
to the whole real axis, we can exploit the infinitesimal quantity $\pm i \delta$
to get rid of the boundary term at infinity. For $\gamma >0$, it can be easily
checked that the function
\begin{eqnarray}
f_{m}^{R,A}(E)= \mp i
\int \!\! dt
\frac{ \theta\left(\pm t \right) }{\cos (\sqrt{\gamma} t )} \,
 e^{-i(E+\eta/\gamma) \tau(t)} \nonumber \\
\times e^{i(\tilde{\omega}_{m}+\eta/\gamma \pm i \delta)t}
\label{intpos}
\end{eqnarray}
 with the function $\tau (t)$ given by
\begin{eqnarray}
\tau(t) = \frac{1}{\sqrt{\gamma}} \tan \left ( \sqrt{\gamma} t \right),
\end{eqnarray}
 is a solution of Eq. (\ref{eqf}). Here integral (\ref{intpos}) is defined in the
sense of Cauchy principal value for the points $\sqrt{\gamma} t = \pi/2+n \pi$.
This provides the exact result~(\ref{resultath}) for the vortex Green's function of an
arbitrary quadratic potential in the $\omega_c\to \infty$ limit.

\section{Checking the vortex formalism: Case of a 2D parabolic confining potential}
\label{AppendixE}

Recovering the set of two discrete quantum numbers for the circular confinement
potential from the use of a basis of states which is characterized by both
discrete and continuous quantum numbers appears in principle to be a very
challenging task. We show that the quantization of the confining potential
appears in the vortex Green's function formalism through a rather different way
from the usual derivation in the wave function formalism.

\subsection{Standard derivation}

To benchmark our results for the Green's functions, we shall compare the general
expression derived in Sec.~\ref{Section4} from the use of the vortex states formalism with the
exact solution for a circular confining potential. The potential profile given by
\begin{equation}
V({\bf r})=\frac{1}{2} m^{\ast} \omega^{2}_{0} \left( x^{2}+y^{2}\right)
\end{equation}
 leads in a homogeneous magnetic field to the well-known Fock-Darwin spectrum
\begin{equation}
E_{nl}= \hbar \Omega \left(n+ \frac{|l|+1}{2} \right) -\frac{l}{2} \hbar \omega_{c} ,
\label{spectrum}
\end{equation}
where $n=0,1, 2,...$ is a positive integer and $l=0,\pm 1,\pm 2, ...$ a positive
or negative integer. Here $\Omega=\sqrt{\omega_{c}^{2}+4 \omega_{0}^{2}}$ is the
renormalized cyclotron pulsation. The normalized wave functions associated with
the energy spectrum (\ref{spectrum}) are written in polar coordinates ${\bf
r}=(r,\theta)$,
\begin{eqnarray}
\Psi_{n,l}({\bf r})= \frac{1}{L}\sqrt{\frac{n!}{(n+|l|)!} }\,\left(\frac{r}{\sqrt{2}L} \right)^{|l|}
 \, L_{n}^{|l|} \left(\frac{r^{2}}{2L^{2}} \right) e^{-\frac{r^{2}}{4 L^{2}}} \nonumber \\
 \times
\frac{ e^{il \theta} }{\sqrt{2 \pi}}
\label{Fock}
, \hspace*{0.5cm}
\end{eqnarray}
where $L_{n}^{|l|}(z)$ corresponds to the generalized Laguerre polynomial of
degree $n$, and $L=\sqrt{\hbar/m^{\ast}\Omega}$ is the renormalized magnetic
length.

The local density can be directly calculated from the knowledge of the energy
spectrum and the exact wave functions, and is given by
\begin{eqnarray}
n({\bf r}) =\sum_{n=0}^{+\infty} \sum_{l=-\infty}^{+ \infty} n_{F}(E_{nl}) \left|\Psi_{nl}({\bf r}) \right|^{2}.
\end{eqnarray}

The method of projection onto a given Landau level is again obtained by
considering $\omega_{c} \gg \omega_{0}$, keeping terms of order $\omega_{0}/\omega_c$.
This is equivalent to taking $\omega_{c}\to\infty$ with $l_B$ finite.
We thus have $\Omega \approx \omega_{c} +2 \omega_{0}^{2}/\omega_{c}$ and $L \approx l_{B}$,
so that the energy spectrum becomes
\begin{eqnarray}
E_{nl} \approx \hbar \omega_{c} \left( m+\frac{1}{2} \right)+\hbar \omega_{0}
\frac{2\omega_{0}}{\omega_{c}} \left(m+\frac{l+1}{2} \right)
,
\label{spectrumcirc2D}
\vspace*{0.5cm}
\end{eqnarray}
with $m= [n+(|l|-l)/2]\geq 0$ the Landau level index.
According to the second term in the right-hand side of Eq. (\ref{spectrumcirc2D}), the Landau levels are generally nondegenerate as a result of the circular confining potential
characterized by the frequency $\omega_{0} \ll \omega_{c}$.

If we restrict ourselves to the lowest Landau level contribution to the local
density for the sake of simplicity and consider the absence of Landau level
mixing, the exact local density gets simplified into
\begin{eqnarray}
n({\bf r})=\frac{1}{2 \pi l_{B}^{2}}\sum_{l=0}^{+\infty}
n_{F}\left(\frac{\hbar \omega_{c}}{2} + (l+1)\frac{\hbar\omega_{0}^2}{\omega_{c}} \right)
\frac{1}{l!} \left( \frac{r^{2}}{2l_{B}^{2}}\right)^{l} \, \nonumber \\
\times e^{-\frac{r^{2}}{2 l_{B}^{2}}}. \hspace*{0.5cm}
\label{exactcircular}
\end{eqnarray}

\subsection{Derivation within the vortex formalism}

The different parameters for the circular confining potential are $\zeta=
l_{B}^{2} m^{\ast} \omega_{0}^{2}=\hbar \omega_{0}^{2}/\omega_{c} $,
$\gamma=\zeta^{2}/4$, and $\eta({\bf r})=l_{B}^{2} \zeta \left|{\bm
\nabla}V({\bf r}) \right|^{2}/8$. Using these values and the general formula for
the local density (\ref{finaldensity}) obtained from the vortex formalism, we
get for the lowest Landau level contribution ($m=0$) to the local density
\begin{widetext}
\begin{eqnarray}
n({\bf r})
&= &
\frac{1}{2 \pi l_{B}^{2}} \left\{
\frac{1}{2}
+\mathrm{Im}
\int_{0}^{+\infty} \!\!\!\!\!\! dt \, \frac{T}{\sinh\left[\pi T t \right]} \exp
\left[i\left(\mu -\frac{\hbar \omega_{c}}{2}-\frac{\hbar \omega_{0}^{2}}{\omega_{c}}\right)t+\frac{r^{2}}{2 l_{B}^{2}}
\, e^{-i t \hbar \omega_{0}^{2}/\omega_{c}} \right] \, e^{-\frac{r^{2}}{2 l_{B}^{2}}} \right\}
\\
&=&
\frac{1}{2 \pi l_{B}^{2}} \left\{
\frac{1}{2}
+\mathrm{Im}
\int_{0}^{+\infty} \!\!\!\!\!\! dt \, \frac{T}{\sinh\left[\pi T t \right]}
\exp \left[i\left(\mu -\frac{\hbar \omega_{c}}{2}-\frac{\hbar \omega_{0}^{2}}{\omega_{c}}\right)t\right]
\sum_{l=0}^{+\infty}
\frac{1}{l!} \left(\frac{r^{2}}{2 l_{B}^{2}} \right)^{l}
\, e^{-i l t \hbar \omega_{0}^{2}/\omega_{c}} \, e^{-\frac{r^{2}}{2 l_{B}^{2}}} \right\}
\\
&=&
\frac{1}{2 \pi l_{B}^{2}} \left\{
\frac{1}{2}
+
\sum_{l=0}^{+\infty}
\frac{1}{l!} \left(\frac{r^{2}}{2 l_{B}^{2}} \right)^{l} \, e^{-\frac{r^{2}}{2 l_{B}^{2}}}
\int_{0}^{+\infty} \!\!\!\!\!\! dt \, \frac{T}{\sinh\left[\pi T t \right]}
\sin \left[\left(\mu -\frac{\hbar \omega_{c}}{2}-(l+1)\frac{\hbar \omega_{0}^{2}}{\omega_{c}} \right)t\right]
 \right\}
.
\label{densitycircular}
\end{eqnarray}
\end{widetext}
Using the integral
\begin{equation}
\int_{0}^{+\infty} \!\!\!\!\!\! dx \, \frac{\sin (ax)}{ \sinh (b x)}=\frac{\pi}{2 b} \tanh\left(\frac{\pi a}{2 b} \right),
\end{equation}
local density (\ref{densitycircular}) is rewritten as

\begin{eqnarray}
n({\bf r})
=
\frac{1}{2 \pi l_{B}^{2}} \left\{
\frac{1}{2}
+\frac{1}{2}
\sum_{l=0}^{+\infty}
\frac{1}{l!} \left(\frac{r^{2}}{2 l_{B}^{2}} \right)^{l} \, e^{-\frac{r^{2}}{2 l_{B}^{2}}} \right. \nonumber \hspace*{1cm}
\\
\left.
\hspace*{1.5cm} \times \tanh \left[\left(\mu -\frac{\hbar \omega_{c}}{2}-(l+1)\frac{\hbar \omega_{0}^{2}}{\omega_{c}} \right)t\right]
 \right\}
. \hspace*{0.5cm}
\label{presque}
\end{eqnarray}
Finally, by noting that the first term in the right hand side of Eq. (\ref{presque}) can be written as
\begin{eqnarray}
\frac{1}{2}= \frac{1}{2}\sum_{l=0}^{+\infty}\frac{1}{l!} \left(\frac{r^{2}}{2 l_{B}^{2}} \right)^{l} \, e^{-\frac{r^{2}}{2 l_{B}^{2}}}
,
\end{eqnarray}
we arrive at formula (\ref{exactcircular}) for the local density. This
establishes the exact equivalence of general formula (\ref{finaldensity}) and of
Eq. (\ref{exactcircular}) in the particular case of a circularly symmetric confining
potential.


\begin{thebibliography}{99}
\bibitem{note}
To be precise, an overcomplete representation is composed by a set of states that contains subsets which are themselves complete. Therefore, the continuous set of plane waves is also overcomplete in the Hilbert space which is spanned by a countable basis. However, contrary to the coherent states, the plane wave states commonly used in the scattering formalism are not elements of the Hilbert space, and thus can not be considered as physical states.
\end{thebibliography}
\end{document}